# Network Deployment for Maximal Energy Efficiency in Uplink with Multislope Path Loss

Andrea Pizzo, *Student Member, IEEE*, Daniel Verenzuela, *Student Member, IEEE*, Luca Sanguinetti, *Senior Member, IEEE*, Emil Björnson, *Senior Member, IEEE*

**Abstract**

This work aims to design the uplink (UL) of a cellular network for maximal energy efficiency (EE). Each base station (BS) is randomly deployed within a given area and is equipped with $M$ antennas to serve $K$ user equipments (UEs). A multislope (distance-dependent) path loss model is considered and linear processing is used, under the assumption that channel state information is acquired by using pilot sequences (reused across the network). Within this setting, a lower bound on the UL spectral efficiency and a realistic circuit power consumption model are used to evaluate the network EE. Numerical results are first used to compute the optimal BS density and pilot reuse factor for a Massive MIMO network with three different detection schemes, namely, maximum ratio combining, zero-forcing (ZF) and multicell minimum mean-squared error. The numerical analysis shows that the EE is a unimodal function of BS density and achieves its maximum for a relatively small density of BS, irrespective of the employed detection scheme. This is in contrast to the single-slope (distance-independent) path loss model, for which the EE is a monotonic non-decreasing function of BS density. Then, we concentrate on ZF and use stochastic geometry to compute a new lower bound on the spectral efficiency, which is then used to optimize, for a given BS density, the pilot reuse factor, number of BS antennas and UEs. Closed-form expressions are computed from which valuable insights into the interplay between optimization variables, hardware characteristics, and propagation environment are obtained.

## I. Introduction

**K**EEPING up with the ever-growing demand for higher data throughput is the major ambition of future cellular networks [1]. An important question is how to evolve communication technologies to deliver higher throughput without prohibitively increasing the power consumption [2]. This calls for new design mechanisms that provide the user equipments (UEs) with high spectral efficiency at moderate energy costs. There is a broad consensus that this wireless capacity growth can only be achieved with a substantial network densification [3] [4]. The main approaches for this densification are twofold: small-cell networks [5]–[7] and Massive MIMO [8]–[12]. The former relies on a massive deployment of small cells that guarantees lower propagation losses. The latter makes use of a massive number of base station (BS) antennas to simultaneously serve a relatively large number of UEs by means of spatial multiplexing. A combination of both has also received a lot of interest in the research literature (e.g., [13], [14]). Despite being potentially effective in increasing spectral efficiency, both solutions tend to increase the power consumed by the network; small cells increase the number of deployed BSs, whereas Massive MIMO requires more hardware per BS. The aim of this work is to design a cellular network from scratch to achieve maximal energy efficiency (EE), without any a priori assumption on the number of BS antennas, UEs, pilot reuse or BS density.

*A. Main literature*

The optimal deployment of cellular networks has received great attention in the literature. The first attempts were based on the simple Wyner model [15] wherein both BSs and UEs are located on a line at fixed positions. Next, more complex 2D symmetric grid-based deployments (e.g., hexagonal lattice) were considered [16]. Both approaches are not suited for modeling and studying networks characterized by a very irregular and dense structure, as envisioned in future cellular networks. To address this problem, advanced mathematical tools based on stochastic geometry have been employed in the last years (e.g., [17]–[19]). Within the stochastic geometry framework, the locations of BSs form a point process in a compact set whose cardinality is a Poisson distributed random variable that is independent among different disjoint sets. The performance of a cellular network can be measured in many different ways such as coverage probability, throughput and EE [6].Earlier works on the design of EE-optimal cellular networks, equipped with multiple antenna BSs, can be found in [9] and [20] where closed-form expressions are derived for a single-cell scenario and numerical results are given for a multicell setting. The EE analysis of a multicell network is developed in [6], [21], [22] by using stochastic geometry. In [21], the optimization is done while satisfying a quality-of-service requirement per UE. In [6], [22], the use of small-cells together with sleeping strategies is proved to be a promising solution for increasing the EE. Generally speaking, small-cells lead to a higher EE but this gain saturates quickly as the density of small cells increases. In [13], it has been shown that further benefits can be achieved by using Massive MIMO.

As the majority of works in the literature, all the aforementioned ones use the standard path loss model where received power decays like $d^{-\alpha}$ over a distance $d$, where $\alpha$ is called the "path loss exponent". This standard path loss model is quite idealized, and in most scenarios $\alpha$ is itself a function of distance, typically an increasing one [23]. For example, three distinct regimes could be easily identified in a practical environment [24]: a distance-independent "near-field" where $\alpha_0 = 0$, a free-space like regime where $\alpha_1 = 2$, and finally some heavily-attenuated regime where $\alpha_2 > 2$.[1] What happens if densification pushes many BSs into the near-field? An answer to this question can be found in [23], [25], [26] (among others), wherein the authors show that the propagation environment and fading distribution play a key role in identifying network operating regimes for which an increase, saturation, or decrease of the throughput is observed as the network densifies. In the extreme case, ultra BS-densification may even lead to zero throughput. Despite all this, multislope path loss models are not frequently used in the analysis of cellular networks because, in general, they make the theoretical analysis more demanding. This work attempts to solve this issue for the EE maximization problem at hand.

*B. Contributions and outline*

We consider a cellular network in which the BSs are independently and uniformly distributed in a given area according to a homogeneous Poisson point process (H-PPP) of intensity $\lambda$. Each BS is equipped with an arbitrary number $M$ of antennas and serves simultaneously $K$ UEs. Statistical channel inversion power-control is employed in the uplink (UL) to achieve a uniform average signal-to-noise ratio (SNR) across all the UEs. A multislope (distance-dependent) path loss model is considered. Three different linear combining schemes, namely, maximum ratio (MR), zero-forcing (ZF) and multicell minimum mean-squared error (M-MMSE), are used under the assumption that channel state information is acquired by using pilots, which are reused across the network with a factor $\zeta$. The EE of the network is computed by using a lower bound on the average UL spectral efficiency (valid for any combining scheme) as well as a polynomial power consumption model, thoroughly developed in [12]. Numerical results are used to evaluate the impact of BS density $\lambda$ and pilot reuse factor $\zeta$ on the EE of a Massive MIMO network such that $M \gg K \gg 1$. The results show that the EE with a multislope path loss is a unimodal[2] function of $\lambda$. Irrespective of the employed detection scheme, the optimal EE is achieved for relatively small values of $\lambda$ and $\zeta$. This is in sharp contrast to [13] where the adoption of a single-slope path loss model leads to the conclusion that densification is always beneficial for EE; the EE is shown to be a monotonic increasing function of $\lambda$ in [13]. The results show also that, although the "optimal" M-MMSE combiner provides the highest EE, the three different schemes behave similarly in terms of EE and area throughput as BS density increases.

---

[1]Such a situation results even with a simple 2-ray ground reflection, with $\alpha_2 = 4$ in that case.

Motivated by the above analysis, we concentrate on ZF and compute a new closed-form lower bound on the average UL SE. This lower bound is used to analytically find in closed-form the EE-optimal network configuration with respect to $M$, $K$ and $\zeta$ while satisfying a signal-to-interference-plus-noise ratio (SINR) constraint. The closed-form expressions reveal the fundamental interplay between the three design parameters, which are also illustrated numerically. It turns out that ZF allows a higher densification of the network while using a smaller pilot reuse factor and achieving a higher EE than with MR. Both schemes employ almost the same optimal number of antennas per BS to approximately serve the same number of UEs, with a ratio $M/K$ between $4$ and $19$ when using ZF and between $4$ and $27$ for MR depending on the SINR constraint. In addition, ZF is characterized by a smoother EE function, which is more robust to system changes and thus makes it a better choice.

Compared to the preliminary version in [27], this work: ($i$) provides the EE analysis for MR, ZF and M-MMSE; ($ii$) is based on a multislope path loss model and aims at showing its impact on EE when the network is densified; ($iii$) gives more details and insights into the effect of network parameters and circuit power model.

The remainder of this paper is organized as follows.[3] Next section introduces basic notation and describes the cellular network with the underlying assumptions and transmission protocols.

---

[2] A function $f(x)$ is unimodal if it is monotonically increasing for $x \leq m$ and decreasing for $x > m$ for some $m \in \mathbb{R}$.

[3] Upper (lower) bold face letters are used for matrices (column vectors). Sans serif fonts are used for mathematical quantities whereas times new roman fonts are used for acronyms/texts. $\mathbf{I}_N$ is the $N \times N$ identity matrix and $\mathbf{0}$ is the zero vector. $(\cdot)^{\mathrm{T}}$, $(\cdot)^*$ and $(\cdot)^{\mathrm{H}}$ are the transpose, conjugate and conjugate transpose operators, respectively. We use $\mathrm{tr}(\cdot)$ to denote the matrix trace operator and $\|\cdot\|$ for the Euclidean norm vector operator. $\lceil x \rfloor$ is the nearest integer projector whereas $\mathbb{P}(A)$ indicates the probability associated with an event $A$. $\mathbb{E}_{\mathbf{n}}\{\cdot\}$ denotes the expectation operator with respect to the random vector $\mathbf{n}$, whereas $\mathbf{n} \sim \mathcal{N}_{\mathbb{C}}(\mathbf{0}, \mathbf{R_n})$ is the shorthand for a circularly-symmetric normal distribution with covariance matrix $\mathbf{R_n}$. We use $\mathbb{R}^n$, $\mathbb{C}^n$, and $\mathbb{N}^n$ to denote the $n$-dimensional real-valued, complex-valued and nonnegative integer-valued vector spaces. We denote $\Gamma(s; x) = \int_x^\infty t^{s-1} e^{-t} \, dt$ the upper incomplete gamma function.

Section III analyzes the EE of MR, ZF, and M-MMSE based on a realistic circuit power model. In Section IV, we consider the ZF scheme and compute a lower bound on the achievable EE, which is then maximized analytically with respect to $M$, $K$ and $\zeta$. The resulting expressions reveal the fundamental interplay between the three design parameters. Numerical results are used in Section V to validate an alternating optimization algorithm, which allows to optimally design the network. Finally, the major conclusions and implications are drawn in Section VI.

## II. NETWORK MODEL AND PROBLEM STATEMENT

We consider the UL of a cellular network wherein the BSs are spatially distributed at locations $\{x_i\} \in \mathbb{R}^2$ within a compact geographic area according to a H-PPP $\Phi_\lambda = \{x_i; i \in \mathbb{N}\} \subset \mathbb{R}^2$ of intensity $\lambda$ [BS/km$^2$]. Let $A$ be the deployment area of interest, the average number of deployed BSs is simply $\mathbb{E}_{\{x_i\}}\{\Phi_\lambda\} = \lambda A$. Each BS has $M$ antennas and serves $K$ single-antenna UEs over a bandwidth of $B_\mathrm{w}$ [MHz]. These $K$ UEs are selected at random from a very large set according to some scheduling algorithm. We assume that each UE is connected to the closest BS such that the coverage area of a BS is its Poisson-Voronoi cell (see Fig. 1). The $K$ UEs are assumed to be uniformly distributed in the Poisson-Voronoi cell. Without loss of generality, we assume that the "typical UE", which is statistically representative for any other UE in the network [28], has an arbitrary index $k$ and is connected to an arbitrary BS $j$. The network operates according to a synchronous time-division-duplex protocol. We denote by $B_\mathrm{c}$ [Hz] and $T_\mathrm{c}$ [s] the coherence bandwidth and time, respectively. Then, the coherence block is composed of $\tau_\mathrm{c} = B_\mathrm{c} T_\mathrm{c}$ [complex samples]. In each coherence block, $\tau_\mathrm{p}$ samples are used for acquiring channel state information by means of UL pilot sequences, whereas $\tau_\mathrm{u}$ and $\tau_\mathrm{d}$ samples (such that $\tau_\mathrm{c} = \tau_\mathrm{p} + \tau_\mathrm{u} + \tau_\mathrm{d}$) are used for payload transmission in the UL and downlink (DL), respectively. We assume that $\tau_\mathrm{p} = \zeta K$ with $\zeta \geq 1$ being the pilot reuse factor and $\tau_\mathrm{u} = \xi(\tau_\mathrm{c} - \zeta K)$ with $\xi \leq 1$ accounting for the UL payload fraction transmission [12].

### A. Received Signal and Power Control Policy

We call $s_{li} \sim \mathcal{N}_\mathbb{C}(0, p_{li})$ the UL payload signal transmitted from UE $i$ of cell $l$ to its serving BS $l$ with power $p_{li} = \mathbb{E}_s\{|s_{li}|^2\}$. The signal $\mathbf{y}_j \in \mathbb{C}^M$ received at BS $j$ is

$$\mathbf{y}_j = \underbrace{\mathbf{h}_{jk}^j s_{jk}}_{\text{desired signal}} + \underbrace{\sum_{i=1, i \neq k}^{K} \mathbf{h}_{ji}^j s_{ji}}_{\text{intra-cell interference}} + \underbrace{\sum_{l \in \Phi_\lambda \setminus \{j\}} \sum_{i=1}^{K} \mathbf{h}_{li}^j s_{li}}_{\text{inter-cell interference}} + \underbrace{\mathbf{n}_j}_{\text{noise}} \qquad (1)$$

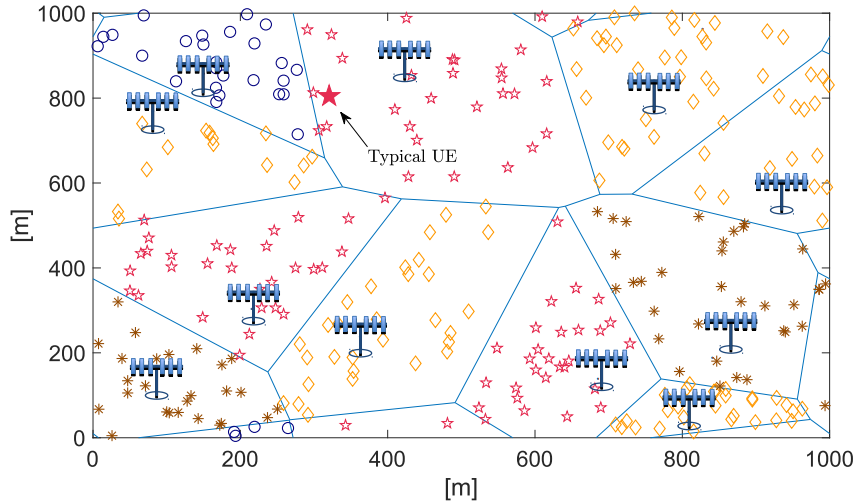

**Fig. 1:** Deployment of a cellular network with BSs drawn from a H-PPP $\Phi_\lambda$, each one serving $K$ randomly located UEs. The "typical" UE $k$ in cell $j$ is highlighted. The UEs of different cells sharing the same pilot subset $\Phi_l$ are depicted with the same marker and color. We consider $A = 1\,\text{km}^2$, $\lambda = 16$, $\zeta = 4$ and $K = 30$.

where $\mathbf{n}_j \sim \mathcal{N}_\mathbb{C}(\mathbf{0}, \sigma^2 \mathbf{I}_M)$ is the additive Gaussian noise, $\mathbf{h}_{li}^j \in \mathbb{C}^M$ is the channel response between UE $i$ in cell $l$ and BS $j$ modeled as uncorrelated Rayleigh fading [11], i.e., $\mathbf{h}_{li}^j \sim \mathcal{N}_\mathbb{C}(\mathbf{0}, \beta_{li}^j \mathbf{I}_M)$, where $\beta_{li}^j$ is the large-scale fading coefficient. We call $d_{li}^j$ the distance of UE $i$ in cell $l$ from BS $j$ and compute $\beta_{li}^j$ according to a general multislope path loss model, which is given by:

$$\beta_{li}^j = \Upsilon_n \, (d_{li}^j)^{-\alpha_n} \qquad (2)$$

with $d_{li}^j \in [R_{n-1}, R_n)$ [km], for $n = 1, \ldots, N$. The coefficients $\{\Upsilon_n\}, \{\alpha_n\}$ are design parameters. Specifically, $0 \leq \alpha_1 \leq \cdots \leq \alpha_N$ are the power decay factors, $0 = R_0 < \cdots < R_N = \infty$ denote the distances at which a change in the power decadence occurs. Setting $N = 1$ yields the widely used single-slope path loss model $\beta_{li}^j = \Upsilon_1 (d_{li}^j)^{-\alpha_1}$.

Following [29], we assume the UEs use a statistical channel inversion power-control policy such that $p_{li} = P_0/\beta_{li}^l$ where $P_0$ is a design parameter. This ensures a uniform ergodic per-antenna received SNR at BS $l$ to all the UEs, which it is given by $\mathbb{E}\{\|\mathbf{h}_{li}^l\|^2 p_{li}\}/(M\sigma^2) = P_0/\sigma^2 = \mathsf{SNR}_0$ and it is assumed to be constant over the coherence block.

## B. Pilot Reuse Policy and Channel Estimation

We assume that a pilot book $\mathbf{\Phi} \in \mathbb{C}^{\tau_\mathrm{p} \times \tau_\mathrm{p}}$ of $\tau_\mathrm{p}$ mutually orthogonal UL sequences is used for channel estimation and call $\boldsymbol{\phi}_{jk} \in \mathbb{C}^{\tau_\mathrm{p}}$ the pilot sequence assigned to the typical UE $k$ in cell $j$. It is assumed to have normalized UL pilot sequences, to obtain a constant power level, and this implies that $\|\boldsymbol{\phi}_{jk}\|^2 = 1$. To avoid cumbersome pilot coordination, we assume that in each coherence block each BS $l$ picks a subset of $K$ different sequences from $\mathbf{\Phi}$, uniformly at random and distribute them among its served UEs. Since $\tau_\mathrm{p} = \zeta K$, we have that the reuse factor is $\zeta = \tau_\mathrm{p}/K > 1$. In other words, there are on average $\mathbb{E}\{\Phi_\lambda\}/\zeta$ cells in the network that share the same pilot subset. This is modeled in each cell through a Bernoulli stochastic variable $a_{l'l} \sim \mathcal{B}(1/\zeta)$ for $l' \neq l$ and $a_{ll} = 1$. Specifically, if $a_{l'l} = 1$ all the UEs in cell $l'$ use the same pilot subset of those in cell $l$ and thus causes pilot contamination [12]. This occurs with probability $\mathbb{P}(a_{l'l} = 1) = 1/\zeta$. Similarly, $a_{l'l} = 0$ indicates that there is no pilot contamination from cell $l'$ to cell $l$ and vice versa, and happens with probability $\mathbb{P}(a_{l'l} = 0) = 1 - 1/\zeta$. To facilitate understanding, Fig. 1 illustrates a pilot allocation snapshot where different markers and colors identify different pilot subsets $\{\Phi_l\}$. We call $\mathbf{Y}_j^\mathrm{p} \in \mathbb{C}^{M \times \tau_\mathrm{p}}$ the signal received at BS $j$ during pilot transmission. The vector $\mathbf{y}_{jli}^\mathrm{p} = \mathbf{Y}_j^\mathrm{p} \boldsymbol{\phi}_{li}^*$ obtained by correlating $\mathbf{Y}_j^\mathrm{p}$ with $\boldsymbol{\phi}_{li}$ takes the form:

$$\mathbf{y}_{jli}^\mathrm{p} = \underbrace{\sqrt{\rho}\sqrt{p_{li}}\mathbf{h}_{li}^j}_{\text{desired pilot}} + \underbrace{\sum_{l' \in \Phi_\lambda \setminus \{l\}} a_{l'l} \sqrt{\rho}\sqrt{p_{l'i}} \mathbf{h}_{l'i}^j}_{\text{interfering pilots}} + \underbrace{\mathbf{N}_j^\mathrm{p} \boldsymbol{\phi}_{li}^*}_{\text{noise}} \qquad (3)$$

where $\mathbf{N}_j^\mathrm{p} \boldsymbol{\phi}_{li}^* \sim \mathcal{N}_\mathbb{C}(\mathbf{0}, \sigma^2 \mathbf{I}_M)$ and the power of the UL transmitted payload signal is scaled by a factor $\rho = P_\mathrm{p}/P_0 \geq 1$ to compensate for the lack of beamforming gain during channel acquisition.

**Corollary 1** (e.g. [12]). *By using $p_{li} = P_\mathrm{p}/\beta_{li}^l$, the MMSE estimate of $\mathbf{h}_{li}^j$ at BS $j$ based on $\mathbf{y}_{jli}^\mathrm{p}$ is*

$$\hat{\mathbf{h}}_{li}^j = \frac{\frac{\beta_{li}^j}{\sqrt{\beta_{li}^l P_\mathrm{p}}}}{\frac{\beta_{li}^j}{\beta_{li}^l} + \sum_{l' \in \Phi_\lambda \setminus \{l\}} a_{l'l} \frac{\beta_{l'i}^j}{\beta_{l'i}^{l'}} + \frac{1}{\mathsf{SNR}_\mathrm{p}}} \mathbf{y}_{jli}^p \qquad (4)$$

*with $\mathsf{SNR}_\mathrm{p} = \rho\, \mathsf{SNR}_0$. The MMSE estimate $\hat{\mathbf{h}}_{li}^j$ and error $\tilde{\mathbf{h}}_{li}^j$, conditioned on a realization of $a_{l'l}$ for all $l', l \in \Phi_\lambda$, are independent and distributed as $\hat{\mathbf{h}}_{li}^j \sim \mathcal{N}_\mathbb{C}(\mathbf{0}, \gamma_{li}^j \mathbf{I}_M)$ and $\tilde{\mathbf{h}}_{li}^j \sim \mathcal{N}_\mathbb{C}(\mathbf{0}, (\beta_{li}^j - \gamma_{li}^j)\mathbf{I}_M)$*

*where*

$$\gamma_{li}^j = \frac{\beta_{li}^j}{\frac{\beta_{li}^j}{\beta_{li}^l} + \sum_{l' \in \Phi_\lambda \setminus \{l\}} a_{l'l} \frac{\beta_{l'i}^j}{\beta_{l'i}^{l'}} + \frac{1}{\mathsf{SNR}_\mathrm{p}}} \frac{\beta_{li}^j}{\beta_{li}^l}. \qquad (5)$$

For notational convenience, we define the collecting of all the estimates in (4) from all UEs in cell $l$ to BS $j$ as $\hat{\mathbf{H}}_l^j = [\hat{\mathbf{h}}_{l1}^j \ldots \hat{\mathbf{h}}_{lK}^j] \in \mathbb{C}^{M \times K}$. Note that the estimate $\hat{\mathbf{h}}_{l'i}^j$ of a UE $i$ in cell $l'$ using the same pilot sequence of UE $i$ in cell $l$ (i.e. $a_{l'l} = 1$) can be obtained from $\hat{\mathbf{h}}_{li}^j$ in (4) as

$$\hat{\mathbf{h}}_{l'i}^j = \sqrt{\frac{\beta_{li}^l}{\beta_{l'i}^{l'}}} \frac{\beta_{l'i}^j}{\beta_{li}^j} \hat{\mathbf{h}}_{li}^j \qquad (6)$$

where $\hat{\mathbf{h}}_{l'i}^j \sim \mathcal{N}_\mathbb{C}(\mathbf{0}, \gamma_{l'i}^j \mathbf{I}_M)$ has variance

$$\gamma_{l'i}^j = \frac{\beta_{li}^l}{\beta_{l'i}^{l'}} \left(\frac{\beta_{l'i}^j}{\beta_{li}^j}\right)^2 \gamma_{li}^j \qquad (7)$$

and estimation error $\tilde{\mathbf{h}}_{l'i}^j \sim \mathcal{N}_\mathbb{C}(\mathbf{0}, (\beta_{l'i}^j - \gamma_{l'i}^j)\mathbf{I}_M)$. The expression in (6) is responsible of pilot contamination with spatially uncorrelated channels; the inability of BS $j$ to separate UEs that use the same pilot [8].

## III. ENERGY EFFICIENCY ANALYSIS

The EE is defined as the amount of information reliably transmitted per unit of energy[4], which is mathematically expressed as [20]:

$$\begin{aligned} \mathsf{EE} &= \frac{\text{Area throughput } [\mathrm{bit/s/km^2}]}{\text{Area power consumption } [\mathrm{W/km^2}]} \\ &= \frac{B_\mathrm{w}[\mathrm{Hz}] \cdot \mathsf{ASE} \, [\mathrm{bit/s/Hz/km^2}]}{\mathsf{APC} \, [\mathrm{W/km^2}]} \end{aligned} \qquad (8)$$

which is measured in [bit/Joule] and can be seen as a benefit-cost ratio, where the service quality (area throughput) is compared with the associated cost (area power consumption). In (8), ASE and APC are the area spectral efficiency (ASE) and area power consumption (APC), respectively.

---

[4]A considerable number of papers on EE analysis has considered misleading EE metrics measured in bit/Joule/Hz, instead of bit/Joule. This is pointless since one cannot make the EE bandwidth-independent: the transmit power is divided over the bandwidth while the noise power is proportional to the bandwidth.

## A. Area Spectral Efficiency

Since the "typical UE" is statistically representative for any other UE in the network [28], the ASE is obtained as $\mathsf{ASE} = \lambda K \, \mathsf{SE}$ in $[\text{bit/s/Hz/km}^2]$ where $\mathsf{SE}$ denotes the average UL spectral efficiency of the typical UE $k$ in cell $j$ and is obtained averaging over different UE positions, pilot allocations and channel realizations. The multiplicative factor $K$ accounts for the sum spectral efficiency of all UEs in cell $j$ and $\lambda$ is the BS density per km$^2$. A lower bound on SE, which holds for any combining scheme, UE positions and pilot allocations is as follows.[5]

**Theorem 1** ([12]). *When the channel is obtained through the MMSE estimator in* (4), *the UL average ergodic channel capacity of the typical UE $k$ in cell $j$ is lower bounded by*

$$\mathsf{SE} \geq \mathsf{SE}' = \xi \left(1 - \frac{K\zeta}{\tau_\text{c}}\right) \mathbb{E}_{\{\mathbf{d},\mathbf{h},\mathbf{a}\}} \left\{ \log_2\left(1 + \mathsf{SINR}'\right) \right\} \tag{9}$$

*where* $\mathsf{SINR}'$ *is the instantaneous SINR given by*

$$\mathsf{SINR}' = \frac{p_{jk}|\mathbf{v}_{jk}^\text{H}\hat{\mathbf{h}}_{jk}^j|^2}{\mathbf{v}_{jk}^\text{H}\left(\sum\limits_{\substack{l\in\Phi_\lambda \\ (l,i)\neq(j,k)}}\sum\limits_{i=1}^{K} p_{li}\hat{\mathbf{h}}_{li}^j\hat{\mathbf{h}}_{li}^{j\text{H}} + \sum\limits_{l\in\Phi_\lambda}\sum\limits_{i=1}^{K} p_{li}(\beta_{li}^j - \gamma_{li}^j)\mathbf{I}_M + \sigma^2\mathbf{I}_M\right)\mathbf{v}_{jk}} \tag{10}$$

*and the expectation $\mathbb{E}_{\{\mathbf{d},\mathbf{h},\mathbf{a}\}}\{\cdot\}$ is computed with respect to UE positions, channel realizations and pilot allocations. The pre-log factor accounts for the pilot overhead.*[6]

The optimal $\mathbf{v}_{jk}$ that maximizes (9) is given as follows.

**Corollary 2** ([12]). *The instantaneous UL SINR in* (10) *for a typical UE $k$ in cell $j$ is maximized by*

$$\mathbf{v}_{jk}^{\text{M-MMSE}} = \left(\sum_{l\in\Phi_\lambda}\sum_{i=1}^{K} p_{li}\left(\hat{\mathbf{h}}_{li}^j(\hat{\mathbf{h}}_{li}^j)^\text{H} + (\beta_{li}^j - \gamma_{li}^j)\mathbf{I}_M\right) + \sigma^2\mathbf{I}_M\right)^{-1} p_{jk}\hat{\mathbf{h}}_{jk}^j. \tag{11}$$

*Proof.* The proof can be found in [12] and it is based on the Rayleigh quotient maximization. □

The optimal combining vector in (11) is known as M-MMSE combiner since it can be proved to be the vector $\mathbf{v}_{jk}$ that minimizes the conditional MSE, that is $\mathbb{E}\{s_{jk} - \mathbf{v}_{jk}^\text{H}\mathbf{y}_j \mid \{\hat{\mathbf{H}}_l^j\}, \{a_{l',l}\}\}$.

---

[5]Note that the UL capacity for a network (such as the one under investigation) with imperfect CSI and inter-cell interference modeled as a shot-noise process is not known yet [30]. As a common practice in these circumstances, we resort to a lower bound.

[6]In each coherence block, BS $j$ uses the first $\tau_\text{p}$ samples for acquiring CSI to decode the UL payload in the remaining $\tau_\text{u}$ samples.

**TABLE I**: Network and system parameters.

| Parameter | Value | Parameter | Value |
|---|---|---|---|
| Fixed power: $P_{\text{FIX}}$ | 5 W | Far-field path loss exponent: $\alpha$ | 4 |
| Power for BS Local Oscillator: $P_{\text{LO}}$ | 0.1 W | Coherence block length: $\tau_{\text{c}}$ | 200 samples |
| Power per BS antennas: $P_{\text{BS}}$ | 0.2 W | Propagation loss at 1 km: $\Upsilon$ | $-148.1$ dB |
| Power for antenna at UE: $P_{\text{UE}}$ | 0.1 W | Bandwidth: $B_{\text{w}}$ | 20 MHz |
| Power for data coding: $P_{\text{COD}}$ | 0.01 W/(Gbit/s) | Deployment area: $A$ | 1 km$^2$ |
| Power for backhaul traffic: $P_{\text{BT}}$ | 0.025 W/(Gbit/s) | UL fraction of payload block: $\xi$ | 1/3 |
| Power for data decoding: $P_{\text{DEC}}$ | 0.08 W/(Gbit/s) | Noise variance: $\sigma^2$ | $-94$ dBm |
| BS computational efficiency: $L_{\text{BS}}$ | 750 Gflops/W | Signal-to-noise ratio of payload block: $\mathsf{SNR}_0$ | 5 dB |
| HPA efficiency: $\eta$ | 0.5 | Signal-to-noise ratio of pilot block: $\mathsf{SNR}_{\text{p}}$ | 15 dB |

**TABLE II**: Power consumed by different combining vectors $\mathbf{v}_{jk}$.

| Combining scheme | $P_{\text{LP-c},j}$ |
|---|---|
| M-MMSE | $\frac{3B_{\text{w}}}{\tau_{\text{c}} L_{\text{BS}}} \left( \lambda A \frac{(M^2+3M)K}{2} + (M^2-M)K + \frac{M^3}{3} + 2M + M\zeta K^2(\zeta-1) \right)$ |
| ZF | $\frac{3B_{\text{w}}}{\tau_{\text{c}} L_{\text{BS}}} \left( \frac{3K^2 M}{2} + \frac{KM}{2} + \frac{K^3-K}{3} + \frac{7}{3}K \right)$ |
| MR | $\frac{3B_{\text{w}}}{\tau_{\text{c}} L_{\text{BS}}} \left( \frac{7}{3}K \right)$ |

Despite its optimality, the M-MMSE combiner has not been used much in the research literature. The majority of works make use of single-cell processing schemes such as single-cell MMSE (S-MMSE), regularized ZF (RZF), ZF and MR, which are all suboptimal. Specifically, they can be obtained as approximations and simplifications of the optimal M-MMSE [12]. For example, S-MMSE can be obtained by considering the intra-cell channel estimates $\hat{\mathbf{H}}_j^j$ only, whereas RZF arises by neglecting interference coming from other cells. The MR combiner $\mathbf{V}_j^{\text{MR}} = \hat{\mathbf{H}}_j^j$ is obtained for low SNR values whereas the ZF combiner

$$\mathbf{V}_j^{\text{ZF}} = \hat{\mathbf{H}}_j^j \left( \left(\hat{\mathbf{H}}_j^j\right)^{\text{H}} \hat{\mathbf{H}}_j^j \right)^{-1} \quad (12)$$

arises for high SNR. In the sequel, we consider M-MMSE, ZF and MR. M-MMSE provides the highest spectral efficiency, but using the highest complexity. MR has the lowest complexity, but also the lowest spectral efficiency. Finally, ZF strikes a good balance between spectral efficiency and complexity [12].

*B. Area Power Consumption*

The APC can be expressed as follows:

$$\text{APC} = \lambda \left( \eta^{-1} P_{\text{TX}} + P_{\text{CP}} \right) \qquad [\text{W/km}^2] \tag{13}$$

where $P_{\text{TX}}$ accounts for the average power usage for UL transmission (payload and pilots) in an arbitrary cell $j$ with $\eta \in (0, 1]$ being the high power amplifier (HPA) efficiency whereas $P_{\text{CP}}$ is the power consumed by circuitry and can be computed as in [13], [20]. Both are evaluated next.

**Corollary 3.** *If $\tau_{\text{u}}$ and $\tau_{\text{p}}$ samples are respectively used in UL for data and pilot transmissions, then the average total power used for transmission is*

$$P_{\text{TX}} = \left( \frac{\tau_{\text{u}} + \rho\, \tau_{\text{p}}}{\tau_{\text{c}}} \right) K \mathcal{U} \tag{14}$$

*where*

$$\mathcal{U} = P_0 \sum_{n=1}^{N} \Upsilon_n^{-1} \frac{\Gamma\left( \frac{2+\alpha_n}{2}; R_n \right) - \Gamma\left( \frac{2+\alpha_n}{2}; R_{n+1} \right)}{(\pi \lambda)^{\alpha_n/2}} \tag{15}$$

*with $\Gamma(s; x) = \int_x^\infty t^{s-1} e^{-t}\, dt$ being the upper incomplete gamma function.*

*Proof.* The BS locations are drawn from a H-PPP $\Phi_\lambda$ and the UEs are uniformly distributed in the Poisson-Voronoi cells. Thus, the distance $d_{jk}^j$ from the typical UE $k$ to its serving BS $j$ is Rayleigh distributed as $d_{jk}^j \sim \mathcal{R}(1/\sqrt{2\pi\lambda})$ and its probability density function is given by $f_d = (2\pi\lambda d_{jk}^j) e^{-\pi\lambda(d_{jk}^j)^2}$ for $d_{jk}^j > 0$ [30]. Besides, $\mathbb{E}_s\{|s_{jk}|^2\} = p_{jk}$ with $p_{jk} = P_0/\beta_{jk}^j$ or $p_{jk} = P_{\text{p}}/\beta_{jk}^j$, depending on wether $s_{jk}$ is a payload or pilot signal, respectively. Within a coherence block, each user transmits data symbols for a fraction of $\tau_{\text{u}}/\tau_{\text{c}}$ and pilot symbols for $\tau_{\text{p}}/\tau_{\text{c}}$. This, together with $P_{\text{p}} = \rho P_0$, leads to $P_{\text{TX}} = \frac{\tau_{\text{u}} + \rho \tau_{\text{p}}}{\tau_{\text{c}}} P_0 K \mathbb{E}_d\{1/\beta_{jk}^j\}$. Finally, using the path loss model in (2) we have that $\mathbb{E}_d\{1/\beta_{jk}^j\} = \sum_{n=1}^{N} \Upsilon_n^{-1} \int_{R_{n-1}}^{R_n} y^{\alpha_n} f_d(y)\, dy$ from which (15) follows by using (69) in Appendix B. □

The power needed to run the circuitry of an arbitrary BS $j$ can be modeled as follows [13], [20]

$$P_{\text{CP}} = \underbrace{P_{\text{FIX}}}_{\substack{\text{fixed}\\\text{power}}} + \underbrace{P_{\text{TC}}}_{\substack{\text{transceiver}\\\text{chain}}} + \underbrace{P_{\text{C-BH}}}_{\substack{\text{coding\&}\\\text{backhauling}}} + \underbrace{P_{\text{CE}}}_{\substack{\text{channel}\\\text{estimation}}} + \underbrace{P_{\text{LP}}}_{\substack{\text{linear}\\\text{processing}}} \tag{16}$$

where $P_{\text{FIX}}$ is the power consumed for site-cooling, control signaling and load-independent backhauling, $P_{\text{TC}}$ for the transceiver chain, $P_{\text{C-BH}}$ for coding and load-dependent backhauling

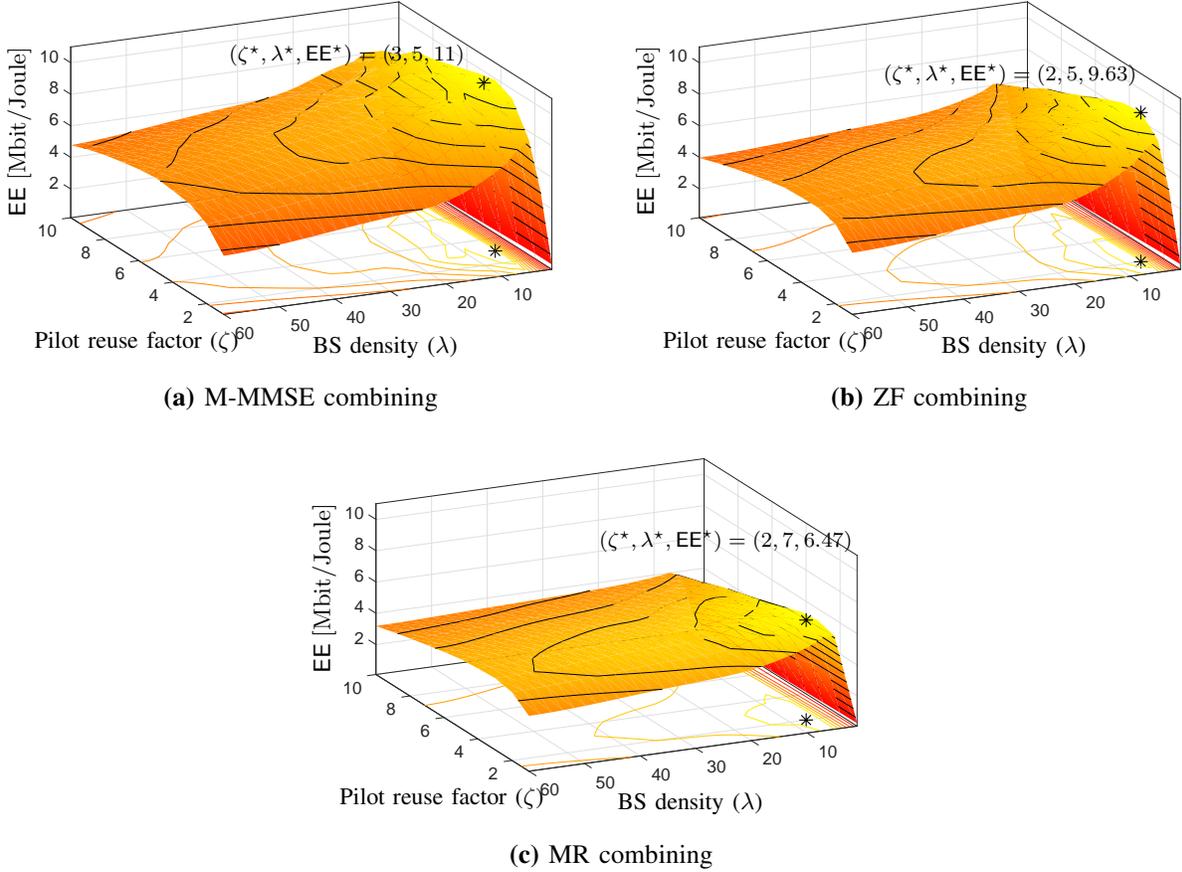

**Fig. 2:** EE (in Mbit/Joule) as a function of $\lambda$ (in BS/km$^2$) and $\zeta$. Results are obtained for MR, ZF and M-MMSE by using Monte Carlo simulations within a Massive MIMO setting with $M = 100$ and $K = 10$. The global optimum is star-marked for which the corresponding values of $\zeta, \lambda$ and EE are also reported.

cost, $P_{\text{CE}}$ for channel estimation process and $P_{\text{LP}}$ for linear processing. All these terms can be expressed as a function of the system parameters reported in Table I. In particular, we have that $P_{\text{TC}} = MP_{\text{BS}} + P_{\text{LO}} + KP_{\text{UE}}$, whereas $P_{\text{C-BH}} = B_{\text{w}} K \, \text{SE} \, (P_{\text{COD}} + P_{\text{DEC}} + P_{\text{BT}})$ with SE that is typically lower bounded, e.g., using (9). The evaluation of $P_{\text{CE}}$ and $P_{\text{LP}}$ requires first to evaluate the computational complexity of channel estimation and linear processing in terms of flops per coherence block.[7] The MMSE estimation has complexity $K(M\tau_{\text{p}} + M)$ since, for any UE, it requires $M\tau_{\text{p}}$ operations to compute $\mathbf{y}_{lij}^p = \mathbf{Y}_j^p \boldsymbol{\phi}_{li}^*$ in (3) and $M$ operations for computing $\hat{\mathbf{h}}_{li}^j$ in (4). To transform these figures into consumed power, we denote with $L_{\text{BS}}$ the computational efficiency of the BS measured in [flops/W] and recall that a complex multiplication requires three

---

[7]For the sake of simplicity, only the number of complex multiplications and divisions is accounted for in the computational complexity analysis. Therefore, a flop accounts only for a real multiplication/division.

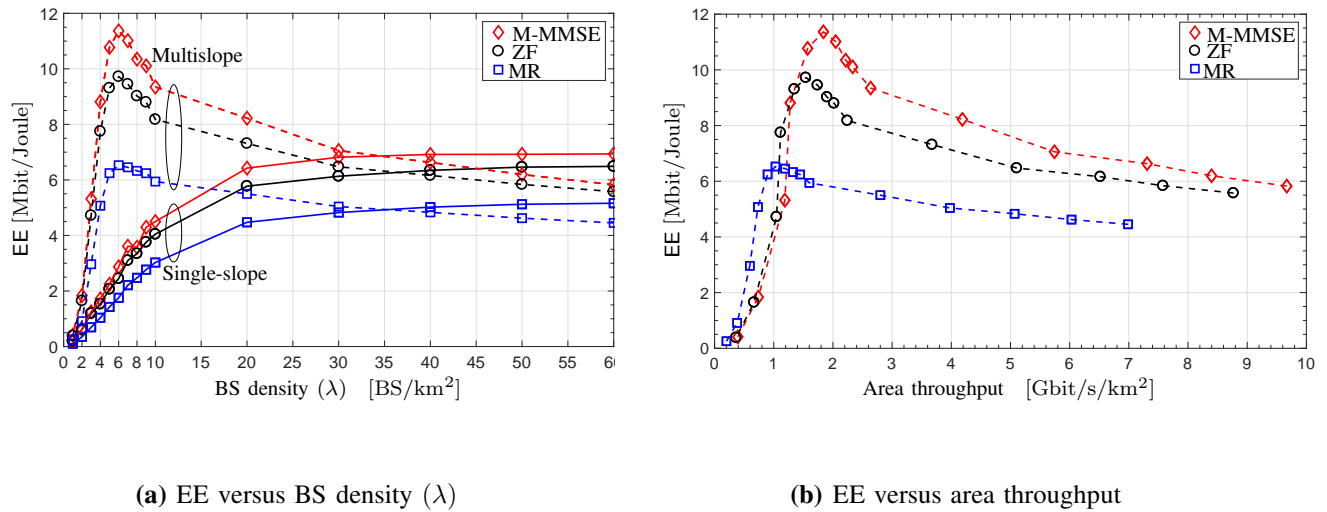

**(a)** EE versus BS density ($\lambda$)

**(b)** EE versus area throughput

**Fig. 3:** EE (in Mbit/Joule) as a function of ($a$) $\lambda$ (in BS/km$^2$) and ($b$) Area Throughput in (in Gbit/s/km$^2$) for the optimal numerical $\zeta^\star$. Results are obtained from Monte Carlo simulations for MR, ZF and M-MMSE combiners within a Massive MIMO setting with $M = 100$ and $K = 10$. In Fig. 3a, both single-slope and multislope path loss models are considered while in Fig. 3b only the multislope is considered.

real multiplications[8]. Thus, since there are $B_\mathrm{w}/\tau_\mathrm{c}$ coherence blocks per second and $\tau_\mathrm{p} = \zeta K$, we obtain

$$P_\mathrm{CE} = \frac{3}{L_\mathrm{BS}} \frac{B_\mathrm{w}}{\tau_\mathrm{c}} KM(\zeta K + 1). \tag{17}$$

The power $P_\mathrm{LP}$ consumed by linear processing can be quantified as $P_\mathrm{LP} = P_\mathrm{LP-r} + P_\mathrm{LP-c}$ where $P_\mathrm{LP-r}$ accounts for the power consumed by reception of payload samples (i.e., evaluation of $y_{jk} = \mathbf{v}_{jk}^\mathrm{H} \mathbf{y}_j$) whereas $P_\mathrm{LP-c}$ is the power required for the computation of the combiner. The former can be quantified as

$$P_\mathrm{LP-r} = \frac{3B_\mathrm{w}}{\tau_\mathrm{c} L_\mathrm{BS}} MK\tau_\mathrm{u} = \frac{3B_\mathrm{w}}{\tau_\mathrm{c} L_\mathrm{BS}} MK\xi\Big(\tau_\mathrm{c} - \zeta K\Big) \tag{18}$$

whereas the latter depends on the $\mathbf{v}_{jk}$ used at BS $j$. Table II provides the power consumed by M-MMSE, ZF and MR [12].

---

[8] Let $x = a + jb$ and $y = c + jd$, then $xy = (ac - bd) + j\left((a + b)(c + d) - ac - bd\right)$ whose computation requires three real multiplications: $ac$, $bd$ and $(a + b)(c + d)$.

*C. Numerical analysis*

We now use the developed power model to design the network for maximal EE. To this end, we adopt the parameters listed in Table I [12].[9] We stress that these parameters tend to be extremely hardware-specific and thus may take substantially different values. The Matlab code that is available online at https://github.com/lucasanguinetti/max-EE-Multislope-Path-Loss enables testing of other values. We consider a deployment area of $A = 1 \mathrm{km}^2$ wherein $\mathbb{E}\{\Phi_\lambda\} = \lambda A$ BSs are randomly deployed according to a H-PPP. A wrap-around topology is used to simulate the H-PPP in the whole $\mathbb{R}^2$ and keep the translation invariance. Inspired by [23] and [14], a bounded $N = 3$ slopes path loss model is used for the large scale fading in (2) with parameters $[\alpha_1, \alpha_2, \alpha_3] = [0, 2, \alpha]$, $[R_1, R_2] = [10, 446]$ [m] and $[\Upsilon_1, \Upsilon_2, \Upsilon_3] = [1, 1, \Upsilon]$ with $\alpha$ and $\Upsilon$ as in Table I. The cut off distance $R_2$ comes from a simple 2-ray ground reflection case, with $R_2 \approx 4 h_\mathrm{t} h_\mathrm{r} / \lambda_\mathrm{c}$ where $h_\mathrm{t} = 10$ and $h_\mathrm{r} = 1.65$ are the antenna heights of BS and UE, and $\lambda_\mathrm{c} = c/f_\mathrm{c}$ is the operating wavelength at a carrier frequency $f_\mathrm{c} = 2\,\mathrm{GHz}$. A Massive MIMO setup is considered, which is roughly characterized by an antenna-UE ratio $M/K = 10$ in order to meet the channel hardening and favorable propagation conditions [12]. Within this setup, we set $M = 100$ and $K = 10$.

Fig. 2 shows the EE of MR, ZF and M-MMSE as a function of BS density and pilot reuse factor $\zeta$. In particular, we consider $\lambda \in \mathcal{L} = \{1, 2, \ldots, 10, 20, \ldots, 60\}$ and $\zeta \in \{1, \ldots, 10\}$ with an average $\mathsf{SNR}_0 = 5$ dB. We see that with all schemes the EE is a pseudo-concave function and has a unique global optimizer $(\lambda^\star, \zeta^\star)$ at which maximum $\mathsf{EE}^\star$ Mbit/Joule is achieved. In the remainder, the superscript $^\star$ is used to indicate optimal values. Each point uniquely determines the EE-optimal network deployment configuration for the corresponding scheme. We notice that M-MMSE provides the highest EE, followed by ZF, while MR achieves the lowest EE. M-MMSE does not only provide the highest EE but has also the smoothest EE around the optimum. This makes it more robust to a variation of network settings (e.g., pilot reuse and BS density). From Fig. 2a, we can see that the maximal EE value with M-MMSE is $\mathsf{EE}^\star = 11$ Mbit/Joule and is achieved at $(\zeta^\star, \lambda^\star) = (3, 5)$. With ZF (see Fig. 2b), we have $\mathsf{EE}^\star = 9.63\,\mathrm{Mbit/Joule}$ at $(\zeta^\star, \lambda^\star) = (2, 5)$, whereas with MR (see Fig. 2c) we obtain $\mathsf{EE}^\star = 6.47\,\mathrm{Mbit/Joule}$ at $(\zeta^\star, \lambda^\star) = (2, 7)$. Note that, irrespective of the combining scheme, the optimal EE is achieved for a pilot reuse factor between $2$ and $3$ and also for a relatively small BS density (e.g., a few BSs per km$^2$). This latter result is in contrast to [13], wherein the EE monotonically increases as $\lambda$ grows large. This was a consequence of the single-slope path loss model adopted in [13]. Further details on this are given next.

---

[9]The interested reader is also referred to the power consumption model developed by IMEC and available on line at the following link http://www.imec.be/powermodel.

Fig. 3a depicts the EE of MR, ZF and M-MMSE as a function of $\lambda \in \mathcal{L}$ for the optimal pilot reuse factors $\zeta^\star$, provided by Fig. 2. Comparisons are made with the EE achieved with a single-slope path loss model with large scale fading parameters $\alpha$ and $\Upsilon$ as in Table I. As expected, in this latter case, the EE is a monotonic non-decreasing function of $\lambda$ for all schemes. For completeness, Fig. 3b illustrates the EE with MR, ZF and M-MMSE as a function of the corresponding average area throughput, obtained for $\lambda \in \mathcal{L}$. We see that there exist operating conditions under which it is possible to jointly increase both the area throughput and EE up to the maximum EE point, but further increases in throughput can only come at a loss in EE. The curves are quite smooth around the maximum EE point; thus, there is a variety of throughput values or, equivalently, BS densities that provide nearly maximum EE with higher area throughput. With M-MMSE, selecting $\lambda = 9 > \lambda^\star = 5$ leads to a $27\%$ increase in area throughput while the EE is reduced by $12\%$ only. The gain is even higher when considering MR and ZF. Specifically, they allow a reduction of respectively $19\%$ and $10\%$ in EE to achieve $46\%$ and $56\%$ higher area throughput.

To summarize, Figs. 2 and 3 show that the additional computational complexity of M-MMSE processing pays off both in terms of EE and area throughput. Moreover, the analysis shows that, for a Massive MIMO setup, reducing the cell size does not bring benefits in terms of EE; the optimal EE is roughly achieved for the same $\lambda$ for all detection schemes.

## IV. Energy Efficiency Maximization with ZF

Monte Carlo simulations were used above to examine the EE of different network configurations for a given pair of $(M, K)$ and detection scheme. In the following, we look at the EE from a different perspective: we design the network from scratch to achieve maximal EE for a given BS density, without any a priori assumption on $M$ and $K$. In doing so, we concentrate on ZF and show how the EE maximization problem can be solved analytically, without the need of heavy Monte Carlo simulations. Moreover, such an analysis exposes fundamental behaviors with respect to network parameters that cannot be easily inferred from the numerical analysis. In particular, we are interested in finding the EE-optimal tuple of parameters $\boldsymbol{\theta} = (\zeta, M, K)$ defined over a set $\Theta = \{\boldsymbol{\theta} : 1 \leq \zeta < \tau_{\mathrm{c}}/K, (M, K) \in \mathbb{N}^2\}$ where the channel coherence block length $\tau_{\mathrm{c}}$ represents the upper limit on the pilot signaling overhead. To this end, we resort to an alternative lower bound for the average ergodic capacity, which is called Use-and-then-Forget (UatF) bound that is as follows.

**Theorem 2** (e.g. [12]). *The UL average ergodic channel capacity of the typical UE $k$ in cell $j$ is lower bounded by* $\mathsf{SE} \geq \xi_{\mathrm{UL}}\left(1 - \frac{K\zeta}{\tau_{\mathrm{c}}}\right)\mathbb{E}_{\mathbf{d}}\left\{\log_2\left(1 + \mathsf{SINR}\right)\right\}$ *where the SINR of the typical UE, conditioned on a realization of UEs locations, is given by*

$$\mathsf{SINR} = \frac{p_{jk}|\mathbb{E}_{\{\mathbf{h},\mathbf{a}\}}\{\mathbf{v}_{jk}^{\mathrm{H}}\mathbf{h}_{jk}^{j}\}|^2}{\sum\limits_{l\in\Phi_\lambda}\sum\limits_{i=1}^{K} p_{li}\mathbb{E}_{\{\mathbf{h},\mathbf{a}\}}\{|\mathbf{v}_{jk}^{\mathrm{H}}\mathbf{h}_{li}^{j}|^2\} - p_{jk}|\mathbb{E}_{\{\mathbf{h},\mathbf{a}\}}\{\mathbf{v}_{jk}^{\mathrm{H}}\mathbf{h}_{jk}^{j}\}|^2 + \sigma^2\mathbb{E}_{\{\mathbf{h},\mathbf{a}\}}\{\|\mathbf{v}_{jk}\|^2\}}. \quad (19)$$

The lower bound in Theorem 2 is less tight than the previous bound in Theorem 1 since the instantaneous channel estimates are not utilized during signal detection [11]. However, it allows to compute a tractable lower bound on the average UL spectral efficiency when ZF is used.

**Lemma 1.** *When the channel is obtained through the MMSE estimator in (4), the UL powers $\{p_{jk}\}$ are chosen as $p_{jk} = P_0/\beta_{jk}^j$ and the ZF combining is chosen as in (12), a lower bound on the UL average ergodic channel capacity of the typical UE $k$ in cell $j$, computed using the UatF bound in (19), is given by* $\underline{\mathsf{SE}} = \xi\left(1 - \frac{K\zeta}{\tau_{\mathrm{c}}}\right)\log_2\left(1 + \underline{\mathsf{SINR}}\right)$ *where*

$$\underline{\mathsf{SINR}} = \frac{M-K}{\underbrace{\mathsf{INT}}_{\text{Interference plus noise}} + \underbrace{(M-K)\mu_2/\zeta}_{\text{Pilot contamination}}} \quad (20)$$

*with*

$$\mathsf{INT} = \left(K + \frac{1}{\mathsf{SNR}_0}\right)\left(1 + \frac{\mu_1}{\zeta} + \frac{1}{\mathsf{SNR}_{\mathrm{p}}}\right) + \frac{K}{\zeta}\left(\mu_1^2 + \mu_2\right) + K\mu_1\left(1 + \frac{1}{\mathsf{SNR}_{\mathrm{p}}}\right) - K\left(1 + \frac{\mu_2}{\zeta}\right) \quad (21)$$

*and $\mu_\kappa$ for $\kappa = 1, 2$*

$$\mu_\kappa = 2\sum_{n=1}^{N}\frac{\Gamma\left(2; \pi\lambda R_{n-1}^2\right) - \Gamma\left(2; \pi\lambda R_n^2\right)}{\kappa\alpha_n - 2} + \frac{2\,c_n(\kappa)}{(\pi\lambda)^{\frac{\kappa\alpha_n}{2}-1}} \\ \left(\Gamma\left(1 + \frac{\kappa\alpha_n}{2}; \pi\lambda R_{n-1}^2\right) - \Gamma\left(1 + \frac{\kappa\alpha_n}{2}; \pi\lambda R_n^2\right)\right) \quad (22)$$

*with $c_n(\kappa) = -\frac{R_n^{2-\kappa\alpha_n}}{\kappa\alpha_n - 2} + \sum_{i=n+1}^{N}\left(\frac{\Upsilon_i}{\Upsilon_n}\right)^\kappa \frac{R_{i-1}^{2-\kappa\alpha_i} - R_i^{2-\kappa\alpha_i}}{\kappa\alpha_i - 2}.$*

*Proof.* The proof is available in the Appendices and is articulated in two parts: the first part is given in Appendix A wherein all the expectations in (19) with respect to channel and pilot realizations are computed, while the second part is given in Appendix B where the expectation with respect to both BS and UE locations is computed. □

Notice that the numerator of SINR in (20) scales with $M - K$ since each BS sacrifices $K$ degrees of freedom for interference suppression within the cell. The pilot contamination term scales also with $M - K$ and accounts for the coherent interference due to UEs that use the same pilot sequence as the typical UE. Many of the interference terms in INT increase with $K$ since having more UEs leads to both more intra-cell and inter-cell interference due to the imperfect CSI and lack of multicell processing. Comparing the above expressions with those obtained in [13] with MR combining and a single-slope path loss model, it follows that the numerator scales with $M$ rather than $M - K$ since MR combining overcome the interference and noise by amplifying the signal of interest using the full array gain of $M$. The interference from other cells is the same for both schemes except for the extra negative term in (21) given by $K(1 + \mu_2/\zeta)$, which makes ZF combining preferable to MR combining whenever the reduced interference is more substantial than the loss in array gain. The pilot contamination term scales also with $M$ (as the useful signal) rather than $M - K$ as in (20) since MR combining does not benefit from interference suppression.

## A. Problem Statement

The EE maximization problem using Lemma 1 is formulated as

$$\boldsymbol{\theta}^\star = \arg\max_{\boldsymbol{\theta} \in \Theta} \underline{\mathsf{EE}}(\boldsymbol{\theta}) = \frac{B_{\mathrm{w}} \underline{\mathsf{ASE}}(\boldsymbol{\theta})}{\underline{\mathsf{APC}}(\boldsymbol{\theta})} \quad (23)$$

$$\text{subject to} \quad \underline{\mathsf{SINR}}(\boldsymbol{\theta}) = \gamma$$

where $\underline{\mathsf{ASE}}(\boldsymbol{\theta}) = \lambda K \underline{\mathsf{SE}}(\boldsymbol{\theta})$ and $\gamma > 0$ is a design parameter, $\underline{\mathsf{SINR}}(\boldsymbol{\theta})$ is given in (20) and $\underline{\mathsf{APC}}(\boldsymbol{\theta})$ can be obtained from (13). This constraint is imposed to avoid that the EE optimization may lead to an optimizer with poor spectral efficiency. More details are given next. By adopting the power consumption model developed in Section III and expanding the contribution due to $P_{\mathrm{TX}}$ in (14) and $P_{\mathrm{CP}}$ in (16)–(18), the PC with ZF combining can be rewritten as:

$$\underline{\mathsf{APC}}(\boldsymbol{\theta}) = \lambda\Big(\mathcal{C}_0 + \mathcal{C}_1 K - \zeta \mathcal{C}_2 K^2 + \mathcal{C}_3 K^3 + \mathcal{D}_0 M + \mathcal{D}_1 MK + \mathcal{D}_2 MK^2\Big) + \mathcal{A}\, B_{\mathrm{w}}\, \underline{\mathsf{ASE}} \quad (24)$$

with $\mathcal{C}_0 = P_{\mathrm{FIX}} + P_{\mathrm{LO}}, \mathcal{C}_1 = P_{\mathrm{UE}} + 5B_{\mathrm{w}}/(\tau_{\mathrm{c}} L_{\mathrm{BS}}) + \mathcal{U}(1 + 1/\tau_{\mathrm{c}})K, \mathcal{C}_2 = \mathcal{U}/\tau_{\mathrm{c}}, \mathcal{C}_3 = B_{\mathrm{w}}/(\tau_{\mathrm{c}} L_{\mathrm{BS}})$, $\mathcal{D}_0 = P_{\mathrm{BS}}, \mathcal{D}_1 = 3B_{\mathrm{w}}(5/2 + \tau_{\mathrm{c}})/(\tau_{\mathrm{c}} L_{\mathrm{BS}}), \mathcal{D}_2 = 9B_{\mathrm{w}}/(2\tau_{\mathrm{c}} L_{\mathrm{BS}})$ and $\mathcal{A} = P_{\mathrm{COD}} + P_{\mathrm{DEC}} + P_{\mathrm{BT}}$. Note that the functional dependence of $\underline{\mathsf{APC}}(\boldsymbol{\theta})$ on $\lambda$ is due to the term $\mathcal{U}$ given in (15), which depends on the transmit power. Due to the unavoidable inter-cell interference in cellular networks, (23) is only feasible for some values of $\gamma$. This feasible range is obtained in [13] observing that SINR is a monotonically increasing function of $M$ and the constraint $\zeta < \tau_c/K$ must be satisfied, which leads to $\gamma < \tau_c/(K\mu_2)$.

## B. Optimal Pilot Reuse Factor

Next, the optimal pilot reuse factor $\zeta^\star$ for problem (23) is computed.

**Lemma 2.** *Consider any pair of $(M, K)$ for which the problem (23) is feasible. The SINR equality constraint in (23) is satisfied by selecting*

$$\zeta^\star(M, K) = \frac{B_1(M, K)\gamma}{M - K - B_2(K)\gamma} \tag{25}$$

*where $B_1 : \mathbb{N}^2 \to \mathbb{R}$ and $B_2 : \mathbb{N} \to \mathbb{R}$ are given by*

$$B_1(M, K) = K\big(\mu_1(1 + \mu_1) - \mu_2\big) + M\mu_2 + \frac{\mu_1}{\mathsf{SNR}_0} \tag{26}$$

$$B_2(K) = K\left(\frac{1}{\mathsf{SNR}_\mathrm{p}} + \mu_1\Big(1 + \frac{1}{\mathsf{SNR}_\mathrm{p}}\Big)\right) + \frac{1 + \frac{1}{\mathsf{SNR}_\mathrm{p}}}{\mathsf{SNR}_0} \tag{27}$$

*Proof.* The optimal $\zeta$ follows from the constraint (23), which parametrizes the solution set with respect to $M$ and $K$. □

The above lemma provides insights into how the EE-optimal pilot reuse factor $\zeta^\star$ depends on the other system parameters. In particular, it shows that $\zeta^\star$ must increase with $K$ to guarantee a certain average SINR equal to $\gamma$. This is intuitive since increasing $\zeta$ leads to better channel estimation which can partially suppress the increased interference due to more UEs. Comparing (25) with the optimal pilot reuse factor

$$\zeta^\star_{\mathrm{MR}}(M, K) = \frac{B_1(M, K)\gamma + 2K\mu_2\gamma}{M - K\gamma - B_2(K)\gamma} \tag{28}$$

obtained in [13] with MR combining, it follows that with MR the denominator scales with $K\gamma$ rather than $K$ and we have a positive extra term in the numerator. Since usually $\gamma \geq 1$ (to ensure reasonable average spectral efficiency constraint), it turns out that a smaller pilot reuse factor can be used with ZF due to its interference suppression capabilities. Notice that $\zeta^\star$ is a decreasing function of $M$, $\mathsf{SNR}_0$ and $\mathsf{SNR}_\mathrm{p}$. This is because all these parameters amplify the desired signal, which, as a consequence, improves the channel estimation and makes the system operate in a less noise limited regime. The pilot reuse factor $\zeta^\star$ reduces as the path loss exponents $\{\alpha_n\}$ increase (since $B_1$ and $B_2$ are reduced), which is natural since inter-cell interference decays more quickly.

**TABLE III:** Optimization parameters.

| Parameter | Value | Parameter | Value |
|---|---|---|---|
| $a_0$ | $\frac{\gamma}{\tau_\text{c}}\mu_2 K^2$ | $b_0$ | $\frac{\gamma}{\tau_\text{c}}\left(\mu_1(1+\mu_1)+\mu_2(\bar{c}-1)\right)$ |
| $a_1$ | $\frac{\gamma K}{\tau_\text{c}}\left(K(\mu_1(1+\mu_1)-\mu_2)+\frac{\mu_1}{\mathsf{SNR}_0}\right)$ | $b_1$ | $\frac{\gamma}{\tau_\text{c}}\mu_1\frac{1}{\mathsf{SNR}_0}$ |
| $a_2$ | $K$ | $b_2$ | $\bar{c}-1-\mu_1\gamma\left(1+\frac{1}{\mathsf{SNR}_\text{p}}\right)-\frac{\gamma}{\mathsf{SNR}_\text{p}}$ |
| $a_3$ | $K\left(1+\gamma\left(\frac{1}{\mathsf{SNR}_\text{p}}+\mu_1\left(1+\frac{1}{\mathsf{SNR}_\text{p}}\right)\right)\right)+\frac{\gamma}{\mathsf{SNR}_0}\left(1+\frac{1}{\mathsf{SNR}_\text{p}}\right)$ | $b_3$ | $\frac{\gamma}{\mathsf{SNR}_0}\left(1+\frac{1}{\mathsf{SNR}_\text{p}}\right)$ |
| $a_4$ | $\mathcal{D}_0 K+\mathcal{D}_1 K^2+\mathcal{D}_2 K^3$ | $b_4$ | $\mathcal{C}_0$ |
| $a_5$ | $\mathcal{C}_0+\mathcal{C}_1 K+\mathcal{C}_3 K^3$ | $b_5$ | $\mathcal{C}_1+\mathcal{D}_0\bar{c}$ |
| $a_6$ | $\tau_\text{c}\mathcal{C}_2 K$ | $b_6$ | $\mathcal{D}_1\bar{c}$ |
| $a_7$ | — | $b_7$ | $\mathcal{C}_3+\mathcal{D}_2\bar{c}$ |
| $a_8$ | — | $b_8$ | $\tau_\text{c}\mathcal{C}_2$ |

## C. Optimal Number of Antennas per BS and Number of UEs

Plugging $\zeta^\star$ as in (25) into (23), the EE maximization problem becomes

$$\begin{aligned}\underset{(M,K)\in\mathbb{N}^2}{\text{maximize}}\quad &\underline{\mathsf{EE}}(\zeta^\star,M,K)=\frac{B_\text{w}\underline{\mathsf{ASE}}(\zeta^\star,M,K)}{\underline{\mathsf{APC}}(\zeta^\star,M,K)}\\ \text{subject to}\quad &1\leq\zeta^\star(M,K)\leq\tau_\text{c}/K.\end{aligned} \quad (29)$$

Next, we look for the optimal values of $M$ and $K$ in the above problem. We start by considering an integer-relaxed version of the original problem (29) obtained by relaxation of the domain set. For analytic tractability, we replace $M$ with $\bar{c}=M/K$, i.e., the number of BS antennas per UE. This yields:

$$\begin{aligned}\underset{(\bar{c},K)\in\mathbb{R}^2}{\text{maximize}}\quad &\underline{\mathsf{EE}}(\zeta^\star,\bar{c},K)=\frac{B_\text{w}\underline{\mathsf{ASE}}(\zeta^\star,\bar{c},K)}{\underline{\mathsf{APC}}(\zeta^\star,\bar{c},K)}\\ \text{subject to}\quad &\frac{K}{\tau_\text{c}}\leq\frac{\zeta^\star(\bar{c},K)K}{\tau_\text{c}}\leq 1\end{aligned} \quad (30)$$

with

$$\frac{\zeta^\star(\bar{c},K)K}{\tau_\text{c}}=K\frac{B_1(\bar{c},K)\gamma/\tau_\text{c}}{K(\bar{c}-1)-B_2(K)\gamma}. \quad (31)$$

However, this is still a non-convex problem that is hard to solve. To this end, we resort to an iterative alternating optimization algorithm in which we optimize over one variable at a time while the other one is kept fixed. By doing so, the objective of (29) turns out being convex in each variable $(M, K)$ and also the descent directions at each iteration can be computed in closed-form. The integer-valued solutions are finally retrieved from the relaxed ones by projection onto $\mathbb{N}^2$.

*1) Optimal Number of Antennas per BS:* We look for the optimal $\bar{c}$ when $K$ is given.

**Lemma 3.** *For any fixed $K > 0$ such that* (30) *is feasible, the EE is maximized by*

$$\bar{c}^\star = \min\left(\max\left(\bar{c}_0, \bar{c}_1\right), \bar{c}_2\right) \tag{32}$$

*with* $\bar{c}_1 = \frac{a_1 + a_3}{a_2 - a_0}$, $\bar{c}_2 = \frac{\frac{K}{\tau_c} a_1 + a_3}{a_2 - \frac{K}{\tau_c} a_0}$, *and*

$$\bar{c}_0 = \frac{r_1}{r_0} + \sqrt{-\frac{q_0}{q_2} - \frac{q_1}{q_2}\frac{r_1}{r_0} + \left(\frac{r_1}{r_0}\right)^2} \tag{33}$$

*where we used* $r_0 = a_2 - a_0$, $r_1 = a_1 + a_3$, $q_0 = a_1 a_6 + a_3 a_5$, $q_1 = a_3 a_4 + a_0 a_6 - a_2 a_5$ *and* $q_2 = a_2 a_4$, *while all the auxiliary parameters* $\{a_i\}$ *are listed in Table III.*

*Proof.* By using the notation introduced in the lemma, we begin with computing the term $\zeta^\star K/\tau_c$ that appears in both ASE and APC. Plugging (26) and (27) into (31) yields $\zeta^\star K/\tau_c = \frac{a_0 \bar{c} + a_1}{a_2 \bar{c} - a_3}$ such that the objective function in (30) reduces to $\text{EE}(\zeta^\star, \bar{c}, K) = \frac{1 - \frac{a_0 \bar{c} + a_1}{a_2 \bar{c} - a_3}}{a_4 \bar{c} + a_5 - a_6 \frac{a_0 \bar{c} + a_1}{a_2 \bar{c} - a_3}}$ which is a quasi-concave function of $\bar{c}$. By taking the first derivative and equating it to zero, we obtain $\bar{c}_0$ in (33), which corresponds to the solution to the unconstrained problem. Further, the constraint in (30) can be rewritten as $1 \leq \frac{a_0 \bar{c} + a_1}{a_2 \bar{c} - a_3} \leq \frac{\tau_c}{K}$ which implies $\bar{c}_1 \leq \bar{c} \leq \bar{c}_2$. This yields the desired result. $\square$

Notice that the constraint $\bar{c} \geq \bar{c}_1$ is active for $\gamma \leq \tau_c/(\mu_2 K)$. Assuming single-slope pathloss model with typical values: $\alpha = 4$, $\tau_c = 200$ symbols and a very high number of UEs (worst case) as $K = 24$, e.g., then we obtain $\gamma \leq 23$, namely a gross spectral efficiency constraint of 4.6 bit/s/Hz/UE, which implies that this constraint is always active for practical cases. The second constraint $\bar{c} \leq \bar{c}_2$ is instead active for $\gamma \leq \tau_c^2/(\mu_2 K^2)$, which obviously applies even more. This lemma shows how the optimal $\bar{c}$ depends on the other system parameters. In particular, we see that $\bar{c}_0$ increases roughly linearly with $K$ and $\gamma$. This is reasonable since the network tends to equip the BSs with more antennas in order to guarantee an increase of the minimum average SINR to each UE. In contrast, the contrary happens with respect to the the circuit power parameters given by $\mathcal{C}_0 = P_{\text{FIX}} + P_{\text{SYN}}$ and $\mathcal{C}_1 = P_{\text{UE}} + 5B_{\text{w}}/(\tau_c L_{\text{BS}}) + \mathcal{U}(1 + 1/\tau_c)$. In particular, $\bar{c}_0$ depends on the BS density as $\lambda^{\alpha_n/4}$ (since $\mathcal{U}$ in $a_5$ is reduced as $\lambda^{-\alpha_n/2}$); larger antenna arrays must be used if the BS density increases. The same happens with respect to $\mathcal{D}_0 = P_{\mathrm{BS}}$ since it becomes more costly to have additional antennas when $P_{\mathrm{BS}}$ increases. Finally, fewer antennas are needed when $\mathsf{SNR}_0$ and $\mathsf{SNR}_{\mathrm{p}}$ are increased since the rate requirement is achieved by using a higher transmitted power during either data transfer or channel estimation.

*2) Optimal Number of UEs per cell:* We now look for the optimal $K$ when $\bar{c}$ is given.

**Lemma 4.** *For any fixed $\bar{c} > 0$ such that the relaxed problem* (30) *is feasible, the optimal number of UEs is*

$$K^\star = \max\left(K_2, \max\left(K_{1,1}, \min\left(K_0, K_{1,2}\right)\right)\right) \tag{34}$$

*where $K_0$ is the real root of the quintic equation $\sum_{i=0}^{5} p_i x^i = 0$ with $p_0 = 2(-m_1 + m_2)$, $p_1 = n_1(-m_1 + m_2) + 3m_3 n_0$, $p_2 = n_1(-m_3 + 3m_1)$, $p_3 = (-m_1 + m_2)n_3 - m_3 n_2$, $p_4 = 2n_4(-m_1 + m_2)$, $p_5 = -m_3 n_4$ and we define $K_2 = -\frac{b_1 + b_3/\tau_{\mathrm{c}}}{b_0 - b_2/\tau_{\mathrm{c}}}$ and*

$$K_{1,1} = \frac{-(b_1 - b_2) - \sqrt{(b_1 - b_2)^2 - 4b_0 b_3}}{2b_0} \tag{35}$$

$$K_{1,2} = \frac{-(b_1 - b_2) + \sqrt{(b_1 - b_2)^2 - 4b_0 b_3}}{2b_0}. \tag{36}$$

*Proof.* As already done in Lemma 3, we begin with computing the term $K\zeta^\star/\tau_{\mathrm{c}}$, which is given by $K \frac{b_0 K + b_1}{b_2 K - b_3}$, being $\{b_i\}$ auxiliary parameters that are listed in Table III. By doing so, the objective function in (30) can be rewritten as

$$\underline{\mathrm{EE}}(\zeta^\star, \bar{c}, K) = \frac{-m_3 K^3 + m_2 K^2 - m_1 K}{n_4 K^4 + n_3 K^3 + n_2 K^2 + n_1 K - n_0} \tag{37}$$

with $m_1 = b_3$, $m_2 = b_2 - b_1$, $m_3 = b_0$ and $n_0 = b_3 b_4$, $n_1 = b_2 b_4 - b_3 b_5$, $n_2 = b_2 b_5 - b_3 b_6 - b_1 b_8$, $n_3 = b_2 b_6 - b_3 b_7 - b_0 b_8$ and $n_4 = b_2 b_7$. Then, this lemma can be proved by taking the derivative of (37) with respect to $K$ and then considering the constraints on $\zeta^\star K/\tau_{\mathrm{c}}$. □

In particular, it is trivial to show that constraint $K_2$ is active if and only if $\gamma \leq \gamma_2^{-1}$ with $\gamma_2 = \left(\mu_1^2 + \mu_1\left(2 + \frac{1}{\mathsf{SNR}_{\mathrm{p}}}\right) + \mu_2(\bar{c} - 1) + \frac{1}{\mathsf{SNR}_{\mathrm{p}}}\right)/(\bar{c} - 1)$. Notice that there exists no generic closed-form root expression for a quintic equation but solutions can be easily found by means of exhaustive search over the domain set. This can be further speeded up by using for example a bisection method over a feasible set. To gain insights into how $K^\star$ is affected by the system parameters, assume that the power consumption required for linear processing due to combining at the BS is negligible, which implies $\mathcal{C}_3 = \mathcal{D}_2 \approx 0$. This is relevant as all these terms essentially decrease with the computational efficiency $L_{\mathrm{BS}}$, which is expected to increase rapidly in the future. If $L_{\text{BS}}$ is very large, we can further neglect other terms due to linear processing and channel estimation, i.e. $\mathcal{D}_1 \approx 0$. For the sake of tractability, we assume also that both SNR and BS density are sufficiently large (since $\mathcal{U}$ reduces as $\lambda^{-\alpha_n/2}$ we have $\mathcal{C}_2 \to 0$ and $\mathsf{SNR}_0 \gg \gamma$). Then, the following result is of interest:

**Corollary 4.** *Consider the optimization problem* (30) *where* $\bar{c} = M/K$ *and* $K$ *are relaxed to be real-valued variables. For any fixed* $\bar{c} > 0$ *such that the relaxed problem is feasible, if we let* $L_{\text{BS}} \to \infty^{10}$, $\lambda \gg 1$ *and* $\mathsf{SNR}_0 \gg \gamma$, *the optimal number of UEs is*

$$K_\infty^\star = \frac{\mathcal{C}_0}{\mathcal{C}_1 + \mathcal{D}_0 \bar{c}} \left( \sqrt{1 + \frac{b_2 - b_1}{b_0} \frac{\mathcal{C}_1 + \mathcal{D}_0 \bar{c}}{\mathcal{C}_0}} - 1 \right). \tag{38}$$

*Proof.* If $\bar{c}$ is given, $\mathcal{C}_2 = \mathcal{C}_3 = \mathcal{D}_1 = \mathcal{D}_2 = 0$ and $b_3 = 0$, then (37) reduces to

$$\underline{\mathsf{EE}}_\infty(\zeta^\star, \bar{c}, K) = K \frac{m_2 - K m_3}{n_1 + K n_2} \tag{39}$$

which is a quasi-concave function whose maximum is achieved for (38). $\square$

The above result coincides with that in [13] and shows that, under the above circumstances, $K^\star$ decreases with $\bar{c}^\star$ as $\sqrt{1/\bar{c}^\star}$. From (38), it is found that $K^\star$ increases with the static energy consumption $\mathcal{C}_0$, while it decreases with $\mathcal{C}_1$ and $\mathcal{D}_0$. The same behavior is observed for the optimal number of BS antennas. Therefore, we may conclude that more BS antennas and UEs per cell can be supported only if the increase in circuit power has a marginal effect on the consumed power. In addition, we note that $K^\star$ is a decreasing function of $\gamma$, since the interference increases as more UEs are served. We later show that when using hardware parameters as in Table I and having reasonable SNR values, the optimal $K$ computed using Corollary 4 achieves practically the same performance of the one in Lemma 4 that requires exhaustive search.

*3) Convergence Analysis of the Alternating Optimization:* To summarize, we first show in Lemma 2 how to compute the optimal $\zeta^\star$ for the original optimization problem (23). This leads to (29) that is later relaxed as in (30), and then solved through the alternating method explicated in Algorithm 1.

**Lemma 5.** *Algorithm 1 converges to the global optimum* $\boldsymbol{\theta}^\star$.

*Proof.* See Appendix C. Results are validated by numerical analysis in the sequel. $\square$

---

[10]Notice that in (38) and (39) the subscript $\infty$ is used to emphasize that the provided expressions is an asymptotic behavior.

**Algorithm 1** Alternating optimization for Problem (29)

```
Set  θ★₀ = (ζ★₀, c̄★₀, K★₀);  0 < ε < 1;  k = 1;
```
**while** $\|\boldsymbol{\theta}^\star_k - \boldsymbol{\theta}^\star_{k-1}\| \geq \varepsilon$ **do**

    Compute $\bar{c}^\star_k$ by using Lemma 3;

    Compute $K^\star_k$ by using Lemma 4 (Corollary 4);

    Compute $M^\star_k = \bar{c}^\star_k K^\star_k$;

    Compute $\zeta^\star_k(M^\star_k, K^\star_k)$ by using Lemma 2;

    Collect $\boldsymbol{\theta}^\star_k = (\zeta^\star_k, M^\star_k, K^\star_k)$;  $k = k+1$;

**end while**

**return** $\boldsymbol{\theta}^\star = (\zeta^\star, \bar{c}^\star, K^\star) = (\zeta^\star_k, \lceil M^\star_k \rceil, \lceil K^\star_k \rceil)$

**TABLE IV:** Optimal network design parameters and performance achieved for $\lambda = 10 \text{ BS/km}^2$ and different $\gamma$ values.

| Combiner | EE★ $\left[\frac{\text{Mbit}}{\text{Joule}}\right]$ | Area throughput★ $\left[\frac{\text{Mbit}}{\text{s km}^2}\right]$ | APC★ $\left[\frac{\text{W}}{\text{km}^2}\right]$ | $M^\star$ | $K^\star$ | $\zeta^\star$ | Cell Reuse $1/\zeta^\star$[%] |
|---|---|---|---|---|---|---|---|
| ZF($\gamma = 1$) | 3.81 | 672 | 176 | 53 | 13 | 3.4 | 28.98 |
| ZF($\gamma = 3$) | 3.66 | 607 | 166 | 53 | 6 | 8.02 | 12.47 |
| ZF($\gamma = 7$) | 2.71 | 453 | 167 | 56 | 3 | 16.34 | 6.12 |
| MR($\gamma = 1$) | 3.58 | 617 | 172 | 52 | 12 | 3.80 | 26.28 |
| MR($\gamma = 3$) | 2.96 | 517 | 174 | 58 | 5 | 8.98 | 11.14 |
| MR($\gamma = 7$) | 2.03 | 446 | 220 | 82 | 3 | 17.08 | 5.86 |

## V. Numerical results

Numerical results are now used to design the network and validate the theoretical analysis done in Section IV with ZF and MMSE channel estimation. The circuit power parameters as well as the channel parameters are taken from [13] and [20] and listed in Table I. As for Section III-C, the Matlab code available online[11] enables testing of other values. The power needed to run the network is computed by using the model developed in Section III-B. We consider a squared deployment area of $1 \text{ km}^2$ with wraparound topology wherein $\mathbb{E}\{\Phi_\lambda\} = \lambda A$ BSs are deployed as described in Section II. The transmission bandwidth is $B_\text{w} = 20 \text{ MHz}$ and each coherence block consists of $\tau_\text{c} = 200$ samples. We assume that $\mathsf{SNR}_\text{p} = 15$ dB and $\mathsf{SNR}_0 = 5$ dB. The path loss model is the same as Section III-C. Fig. 4 plots the EE of the network as a function of $\lambda$ with MR, ZF, and M-MMSE with $M = 100$ and $K = 10$. The curve labelled ZF-LB is obtained by using the closed-form expression of the average spectral efficiency provided in Lemma 1 while MR-LB is obtained by extending the results (not shown for space limitations) of [13, Proposition 1] to the considered multislope path loss model. The curves MR, ZF, and M-MMSE refer to the performance of a network in which the average spectral efficiency is numerically evaluated by using the Theorem 2. Several important observations can be made from the results presented in Fig. 4. Firstly, although there is a gap between the lower and upper bounds, the curves behave exactly the same for any value of $\lambda$. This validates the accuracy of the spectral efficiency expressions provided in Lemma 1 and [13, Proposition 1] (recall that for this latter case results need to be first extended to the considered multislope path loss model). Secondly, in all cases the EE is a unimodal function of $\lambda$ and the optimal deployment is achieved for the same (relatively small) BS density.

---

[11]See https://github.com/lucasanguinetti/max-EE-Multislope-Path-Loss

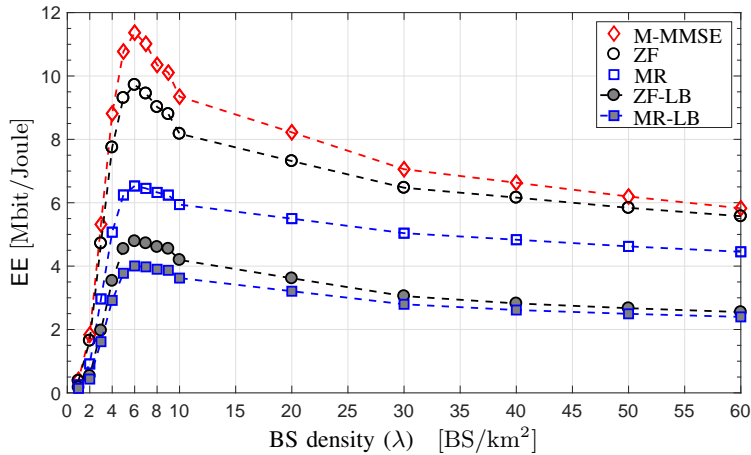

**Fig. 4:** EE (in Mbit/Joule) as a function of $\lambda$ (in BS/km$^2$), for fixed $\mathsf{SNR}_0 = 5$ dB and $\mathsf{SNR}_\mathrm{p} = 15$ dB. Results are obtained for MR, ZF and M-MMSE with $M = 100$ and $K = 10$. The optimal pilot reuse factor $\zeta$ is computed numerically. Curves are obtained by numerically evaluating the bound provided in Theorem 2 (empty-markers) and by using the closed-form expression (filled-markers) in Lemma 1 and in [13, Proposition 1] for MR (extended to the multislope model in (2)).

Fig. 5 shows the EE lower bound as a function of $K$ and $M > K$ with $\gamma = 3$. The pilot reuse is chosen optimally according to (25) and the BS density is fixed to $\lambda = 10\,\mathrm{BS/km^2}$, which is found to be a good compromise between EE and area throughput in Section III. As it is seen, EE is pseudo-concave and has a unique global maximizer (black triangle), which is closely approached by the alternating optimization algorithm in Section IV. The global maximizer is given by the triplet $(M^\star, K^\star, \zeta^\star) = (53, 6, 8)$ and gives a maximum of $\mathsf{EE}^\star = 3.66$ Mbit/Joule. The ratio $M/K$ for the considered setup is roughly 9, which resembles a Massive MIMO setup. The cell reuse $1/\zeta^\star \approx 12\%$ is low enough to ensure robustness against pilot contamination. Compared to [13], the results in Fig. 5 shows that ZF is characterized by a smoother EE function with respect to MR, which makes ZF more robust to system changes and thus a better choice as BS combiner.

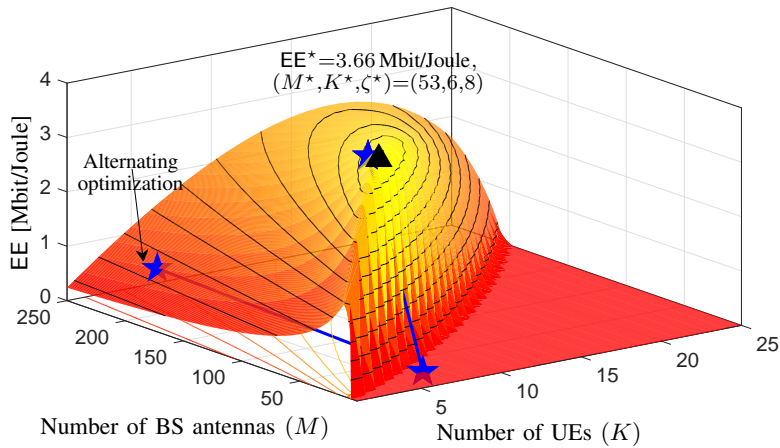

**Fig. 5:** EE (in Mbit/Joule) of ZF as a function of $M$ and $K$, for fixed $\gamma = 3$, $\lambda = 10$ BS/km$^2$, SNR$_0 = 5$ dB and SNR$_\mathrm{p} = 15$ dB. The global optimum obtained from Monte Carlo simulations is indicated with a black triangle. This is compared to the optimum achieved by using Algorithm 1 (blue star).

In Table IV, ZF and MR are compared by using $\lambda = 10$ BS/km$^2$ and with $\gamma \in \{1, 3, 7\}$, which corresponds to the average spectral efficiencies $\log_2(1+\gamma) \in \{1, 2, 3\}$ [bit/s/Hz/UE]. We observe that both schemes require almost the same optimal number of antennas at each BS and serve approximately the same number of UEs. With both schemes, the ratio $M/K$ increases almost linearly with $\gamma$ as stated in Lemma 3. ZF achieves a higher EE than MR. This is a direct consequence of the ZF capabilities to handle intra-cell interference, which allows the BS to serve more UEs with a smaller number of antennas. This has a dual-positive effect: higher area throughput due to multiplexing gain and lower APC because of smaller arrays (especially when the network operates in a high area throughput regime, which requires larger antenna array). Notice also that, as claimed in Lemma 4, $K^\star$ decreases with $\gamma$ since the interference increases as more UEs are served. The pilot reuse factor is slightly smaller than with MR. This happens because ZF mitigates the intra-cell interference and allows a higher inter-cell interference, rather than MR. Nevertheless, we notice that inter-cell interference is not handle by either of them.

## VI. Conclusions

We designed a cellular network for maximal EE with MR, ZF and M-MMSE under the assumption of imperfect CSI and a multislope path loss model. This was formulated as an optimization problem by using a lower bound on the spectral efficiency and a state-of-the-art power consumption model. The variables were pilot reuse factor $\zeta$ and BS density $\lambda$ for a Massive MIMO network. The results showed that the additional computational complexity of M-MMSE processing pays off in terms of EE and area throughput, though all the scheme behaves substantially the same with respect to $\lambda$, that is, reducing the cell size does not bring benefits in terms of EE. To get further insights, we concentrated on ZF and formulated the optimization problem by using stochastic geometry and a new lower bound on the average ergodic spectral efficiency. The variables were pilot reuse factor, number of BS antennas and UEs per BS. The results showed that ZF allows a higher network densification and the use of a smaller pilot reuse factor while achieving a higher EE than with MR combining. Also, it turned out that the EE-optimal configuration resembles a Massive MIMO setup.

## Appendix A — Proof of Lemma 1

To ease understanding, the proof is articulated in two steps. The first aims at computing all the inner expectations in (19) with respect to channel and pilot realizations, whereas the second makes use of all these terms, together with the power allocation policy in Section II, to evaluate the outer expectation with respect to both the BSs' and UEs' locations.

### A.1. Computation of all the expectations in (19)

Let $\mathbf{v}_{jk} = \mathbf{V}_j^{\mathrm{ZF}} \mathbf{e}_k$ be the ZF detector for the typical UE $k$ in cell $j$, with $\mathbf{V}_j^{\mathrm{ZF}}$ as in (12) and $\mathbf{e}_k$ being the $k$th vector of the standard basis.

*Received signal power:* The useful term is computed as:

$$|\mathbb{E}_{\{\mathbf{h},\mathbf{a}\}}\{\mathbf{v}_{jk}^{\mathrm{H}}\mathbf{h}_{jk}^{j}\}|^2 \stackrel{(a)}{=} |\mathbb{E}_{\{\mathbf{h},\mathbf{a}\}}\{\mathbf{v}_{jk}^{\mathrm{H}}\hat{\mathbf{h}}_{jk}^{j}\}|^2 \stackrel{(b)}{=} 1 \tag{40}$$

where $(a)$ follows from Corollary 1 and $(b)$ comes from the ZF combining.

*Noise power:* The noise term is obtained as:

$$\mathbb{E}_{\{\mathbf{h},\mathbf{a}\}}\{\|\mathbf{v}_{jk}\|^2\} \stackrel{(a)}{=} \mathbb{E}_{\{\mathbf{h},\mathbf{a}\}}\left\{\left[((\hat{\mathbf{H}}_j^j)^{\mathrm{H}}\hat{\mathbf{H}}_j^j)^{-1}\right]_{k,k}\right\}$$
$$\stackrel{(a)}{=} \frac{1}{M-K}\mathbb{E}_{\{\mathbf{a}\}}\left\{\frac{1}{\gamma_{jk}^j}\right\} \tag{41}$$

where $(a)$ is due to the ZF combining, $(b)$ exploits both the statistics of $\hat{\mathbf{h}}_{jk}^j$ in Corollary 1, with $\gamma_{jk}^j$ denoted in (5), and the properties of Wishart matrices (e.g., see [9, Proof of Proposition 3] or [31] for a more general treatment).

*Intra-cell interference power:* Consider now the cell of interest $j$ and let us compute the intra-cell interference. Then, for any interferer UE $i$ in cell $j$ we have that

$$\begin{aligned}
&\mathbb{E}_{\{\mathbf{h},\mathbf{a}\}}\{|\mathbf{v}_{jk}^{\text{H}}\mathbf{h}_{ji}^j|^2\} \\
&\stackrel{(a)}{=} \mathbb{E}_{\{\mathbf{h},\mathbf{a}\}}\{\text{tr}(\mathbf{v}_{jk}\mathbf{v}_{jk}^{\text{H}}\hat{\mathbf{h}}_{ji}^j(\hat{\mathbf{h}}_{ji}^j)^{\text{H}}) + \text{tr}(\mathbf{v}_{jk}\mathbf{v}_{jk}^{\text{H}}\tilde{\mathbf{h}}_{ji}^j(\tilde{\mathbf{h}}_{ji}^j)^{\text{H}})\} \\
&\stackrel{(b)}{=} \delta(i-k) + \mathbb{E}_{\{\mathbf{a}\}}\{(\beta_{ji}^j - \gamma_{ji}^j)\mathbb{E}_{\{\mathbf{h}\}}\{\|\mathbf{v}_{jk}\|^2\}\} \\
&\stackrel{(c)}{=} \delta(i-k) + \frac{1}{M-K}\left(\beta_{ji}^j\mathbb{E}_{\{\mathbf{a}\}}\left\{\frac{1}{\gamma_{jk}^j}\right\} - \mathbb{E}_{\{\mathbf{a}\}}\left\{\frac{\gamma_{ji}^j}{\gamma_{jk}^j}\right\}\right)
\end{aligned} \quad (42)$$

with $\delta(\cdot)$ as the delta Kronecker function and where $(a)$–$(b)$ follow from Corollary 1 while using ZF combining for the first term and $(c)$ is due to (41).

*Inter-cell interference power:* The inter-cell interference collected by BS $j$ from cell $l$, e.g., depends on whether the UEs in that cell use the same or a different pilot subset $\Phi_l$ than the one used in cell $j$, that is $(l,i) \in \mathcal{P}_{j,k}$ with $\mathcal{P}_{j,k} = \{(l,i) \in \Phi_\lambda \setminus \{j\} \times \{1,\ldots,K\} : a_{lj} = 1\}$. Consider a UE $i$ that does not cause pilot contamination to UE $k$ of cell $j$. In that case, $(l,i) \notin \mathcal{P}_{j,k}$ and

$$\begin{aligned}
\mathbb{E}_{\{\mathbf{h},\mathbf{a}\}}\{|\mathbf{v}_{jk}^{\text{H}}\mathbf{h}_{li}^j|^2 | a_{lj}=0\} &\stackrel{(a)}{=} \beta_{li}^j \mathbb{E}_{\{\mathbf{h},\mathbf{a}\}}\{\|\mathbf{v}_{jk}\|^2 | a_{lj}=0\} \\
&\stackrel{(b)}{=} \frac{\beta_{li}^j}{M-K}\mathbb{E}_{\{\mathbf{a}\}}\left\{\frac{1}{\gamma_{jk}^j}\bigg| a_{lj}=0\right\}
\end{aligned} \quad (43)$$

where $(a)$ follows from Corollary 1 and the fact that $\mathbf{v}_{jk}$ is a function of $\{\hat{\mathbf{h}}_{ji}^j\}_{i=1}^K$, which are statistically independent from $\{\hat{\mathbf{h}}_{li}^j\}_{i=1}^K$ (and so $\{\mathbf{h}_{ji}^j\}_{i=1}^K$) in presence of no pilot contamination, for which (6) does not hold and $(b)$ is due to (41). Consider an interferer UE $i$ that cause pilot contamination to UE $k$ of cell $j$. Then, $(l,i) \in \mathcal{P}_{j,k}$ and using Corollary 1 leads to $\mathbb{E}_{\{\mathbf{h},\mathbf{a}\}}\{|\mathbf{v}_{jk}^{\text{H}}\mathbf{h}_{li}^j|^2 | a_{lj}=1\} = \mathbb{E}_{\{\mathbf{h},\mathbf{a}\}}\{|\mathbf{v}_{jk}^{\text{H}}\hat{\mathbf{h}}_{li}^j|^2 | a_{lj}=1\} + \mathbb{E}_{\{\mathbf{h},\mathbf{a}\}}\{|\mathbf{v}_{jk}^{\text{H}}\tilde{\mathbf{h}}_{li}^j|^2 | a_{lj}=1\}$. Let us now tackle these contributions to pilot contamination separately. The former is

$$\begin{aligned}
\mathbb{E}_{\{\mathbf{h},\mathbf{a}\}}\{|\mathbf{v}_{jk}^{\text{H}}\hat{\mathbf{h}}_{li}^j|^2 | a_{lj}=1\} &\stackrel{(a)}{=} \frac{(\beta_{li}^j)^2}{\beta_{li}^l \beta_{ji}^j}\mathbb{E}_{\{\mathbf{h},\mathbf{a}\}}\{|\mathbf{v}_{jk}^{\text{H}}\hat{\mathbf{h}}_{ji}^j|^2 | a_{lj}=1\} \\
&\stackrel{(b)}{=} \frac{(\beta_{li}^j)^2}{\beta_{li}^l \beta_{ji}^j}\delta(i-k)
\end{aligned} \quad (44)$$

where in $(a)$ we make use of (6) that accounts for the channel estimates linearity ($a_{lj} = 1$) and $(b)$ follows from the same considerations used for (42) since conditioning has no impact here. The latter term is computed as

$$\mathbb{E}_{\{\mathbf{h},\mathbf{a}\}}\{|\mathbf{v}_{jk}^{\text{H}}\tilde{\mathbf{h}}_{li}^{j}|^2|a_{lj}=1\} \stackrel{(a)}{=} \beta_{li}^{j}\mathbb{E}_{\{\mathbf{h},\mathbf{a}\}}\{\|\mathbf{v}_{jk}\|^2|a_{lj}=1\} - \mathbb{E}_{\{\mathbf{h},\mathbf{a}\}}\{\gamma_{li}^{j}\|\mathbf{v}_{jk}\|^2|a_{lj}=1\}$$

$$\stackrel{(b)}{=} \frac{\beta_{li}^{j}\mathbb{E}_{\{\mathbf{a}\}}\left\{\frac{1}{\gamma_{jk}^{j}}\Big|a_{lj}=1\right\}}{M-K} - \frac{\mathbb{E}_{\{\mathbf{a}\}}\left\{\frac{\gamma_{li}^{j}}{\gamma_{jk}^{j}}\Big|a_{lj}=1\right\}}{M-K}$$

$$\stackrel{(c)}{=} \frac{\beta_{li}^{j}\mathbb{E}_{\{\mathbf{a}\}}\left\{\frac{1}{\gamma_{jk}^{j}}\Big|a_{lj}=1\right\} - \frac{(\beta_{li}^{j})^2}{\beta_{li}^{l}\beta_{ji}^{j}}\mathbb{E}_{\{\mathbf{a}\}}\left\{\frac{\gamma_{ji}^{j}}{\gamma_{jk}^{j}}\Big|a_{lj}=1\right\}}{M-K} \quad (45)$$

where $(a)$ follows from Corollary 1 while keeping the conditioning, $(b)$ is due to (41) and $(c)$ follows from (6). The inter-cell interference term can be computed by considering the probability of having or not pilot contamination between UE $i$ of cell $l$ and the typical UE, i.e., $\mathbb{P}(a_{lj} = 1) = 1/\zeta$ and $\mathbb{P}(a_{lj} = 0) = 1 - 1/\zeta$. Particularly, from (43) – (45) we obtain

$$\mathbb{E}_{\{\mathbf{h},\mathbf{a}\}}\{|\mathbf{v}_{jk}^{\text{H}}\mathbf{h}_{li}^{j}|^2\} = \sum_{p=0}^{1}\mathbb{P}(a_{lj}=p)\,\mathbb{E}_{\{\mathbf{h},\mathbf{a}\}}\{|\mathbf{v}_{jk}^{\text{H}}\mathbf{h}_{li}^{j}|^2|a_{lj}=p\}$$

$$= \frac{1}{\zeta}\frac{(\beta_{li}^{j})^2}{\beta_{li}^{l}\beta_{ji}^{j}}\delta(i-k) - \frac{1}{\zeta}\frac{1}{M-K}\frac{(\beta_{li}^{j})^2}{\beta_{li}^{l}\beta_{ji}^{j}}\mathbb{E}_{\{\mathbf{a}\}}\left\{\frac{\gamma_{ji}^{j}}{\gamma_{jk}^{j}}\Big|a_{lj}=1\right\}$$

$$+ \frac{\beta_{li}^{j}}{M-K}\mathbb{E}_{\{\mathbf{a}\}}\left\{\frac{1}{\gamma_{jk}^{j}}\right\}. \quad (46)$$

*A.2. Computation of the lower bound on the UL SE*

To begin with, let us define the following quantities

$$\vartheta_{ji}^{(1)} = \sum_{l \in \Phi_\lambda \setminus \{j\}} \frac{\beta_{li}^{j}}{\beta_{li}^{l}}, \quad \vartheta_{ji}^{(2)} = \sum_{l \in \Phi_\lambda \setminus \{j\}} \left(\frac{\beta_{li}^{j}}{\beta_{li}^{l}}\right)^2 \quad (47)$$

which are then used to compute the inner expectations in (41), (42) and (46) with respect to the pilot realizations only. Then, the following terms are of interest

$$\mathbb{E}_{\{\mathbf{a}\}}\left\{\frac{1}{\gamma_{jk}^{j}}\right\} = \frac{1}{\beta_{jk}^{j}}\left(1 + \frac{\vartheta_{jk}^{(1)}}{\zeta} + \frac{1}{\text{SNR}_{\text{P}}}\right) \quad (48)$$

$$\mathbb{E}_{\{\mathbf{a}\}}\left\{\frac{\gamma_{ji}^{j}}{\gamma_{jk}^{j}}\right\} = \begin{cases} \mathbb{E}_{\{\mathbf{a}\}}\{1\} = 1 & \text{if } i = k \\ \mathbb{E}_{\{\mathbf{a}\}}\left\{\frac{1}{\gamma_{jk}^{j}}\right\}\mathbb{E}_{\{\mathbf{a}\}}\{\gamma_{ji}^{j}\} \stackrel{(a)}{\geq} \frac{\beta_{ji}^{j}}{\beta_{jk}^{j}}\frac{1+\frac{\vartheta_{jk}^{(1)}}{\zeta}+\frac{1}{\text{SNR}_{\text{P}}}}{1+\frac{\vartheta_{ji}^{(1)}}{\zeta}+\frac{1}{\text{SNR}_{\text{P}}}} & \text{if } i \neq k \end{cases} \quad (49)$$

where (48) follows from (5) with $\mathbb{E}_{\{\mathbf{a}\}}\{a_{lj}\} = 1/\zeta$ and $\vartheta_{jk}^{(1)} \in \mathbb{R}$ that is denoted in (47), while in (49) we exploit the fact that UEs $i$ and $k$ in cell $j$ cannot share the same pilot sequence when $i \neq k$ and $(a)$ comes directly from jointly using Jensen's inequality and (48).[12] The noise term in (19) is computed by plugging (48) into (41) as follows

$$\beta_{jk}^j \mathbb{E}_{\{\mathbf{h},\mathbf{a}\}}\{\|\mathbf{v}_{jk}\|^2\} = \frac{1}{M-K}\left(1 + \frac{\vartheta_{jk}^{(1)}}{\zeta} + \frac{1}{\mathsf{SNR}_{\mathrm{p}}}\right). \qquad (50)$$

The interference contribution is decomposed as the sum of three terms, i.e., the intra-cell interference and the inter-cell interferences due to pilot and no pilot contamination:

$$\sum_{l\in\Phi_\lambda}\sum_{i=1}^{K} \frac{\beta_{jk}^j}{\beta_{li}^l} \mathbb{E}_{\{\mathbf{h},\mathbf{a}\}}\{|\mathbf{v}_{jk}^{\mathrm{H}}\mathbf{h}_{li}^j|^2\} = \underbrace{\sum_{i=1}^{K} \frac{\beta_{jk}^j}{\beta_{ji}^j} \mathbb{E}_{\{\mathbf{h},\mathbf{a}\}}\{|\mathbf{v}_{jk}^{\mathrm{H}}\mathbf{h}_{ji}^j|^2\}}_{\text{intra-cell interference}}$$

$$+ \underbrace{\sum_{l\in\Phi_\lambda\setminus\{j\}}\sum_{i=1}^{K}\frac{\beta_{jk}^j}{\beta_{li}^l}\sum_{p=0}^{1}\mathbb{P}(a_{lj}=p)\,\mathbb{E}_{\{\mathbf{h},\mathbf{a}\}}\{|\mathbf{v}_{jk}^{\mathrm{H}}\mathbf{h}_{li}^j|^2|a_{l,j}=p\}}_{\text{inter-cell interference}}. \qquad (51)$$

The first term accounts for intra-cell interference from all UEs $i$ in cell $j$ when using (48) and (49) into (42) becomes

$$\sum_{i=1}^{K}\frac{\beta_{jk}^j}{\beta_{ji}^j}\mathbb{E}_{\{\mathbf{h},\mathbf{a}\}}\{|\mathbf{v}_{jk}^{\mathrm{H}}\mathbf{h}_{ji}^j|^2\} \leq 1 + \frac{K\left(1+\frac{\vartheta_{jk}^{(1)}}{\zeta}+\frac{1}{\mathsf{SNR}_{\mathrm{p}}}\right) - \sum_{i=1}^{K}\frac{1+\frac{\vartheta_{jk}^{(1)}}{\zeta}+\frac{1}{\mathsf{SNR}_{\mathrm{p}}}}{1+\frac{\vartheta_{ji}^{(1)}}{\zeta}+\frac{1}{\mathsf{SNR}_{\mathrm{p}}}}}{M-K}. \qquad (52)$$

The second term accounts for the interference from all UEs $i$ in cells $l \neq j$ and thus it can be computed substituting (48) – (49) into (46), which reads

$$\sum_{l\in\Phi_\lambda\setminus\{j\}}\sum_{i=1}^{K}\frac{\beta_{jk}^j}{\beta_{li}^l}\mathbb{E}_{\{\mathbf{h},\mathbf{a}\}}\{|\mathbf{v}_{jk}^{\mathrm{H}}\mathbf{h}_{li}^j|^2\} \leq \frac{1}{M-K}\Bigg[\left(1+\frac{\vartheta_{jk}^{(1)}}{\zeta}+\frac{1}{\mathsf{SNR}_{\mathrm{p}}}\right)\sum_{i=1}^{K}\vartheta_{ji}^{(1)} + \frac{(M-K)}{\zeta}\vartheta_{jk}^{(2)}$$
$$-\frac{1}{\zeta}\sum_{i=1}^{K}\vartheta_{ji}^{(2)}\left(\frac{1+\frac{\vartheta_{jk}^{(1)}}{\zeta}+\frac{1}{\mathsf{SNR}_{\mathrm{p}}}}{1+\frac{\vartheta_{ji}^{(1)}}{\zeta}+\frac{1}{\mathsf{SNR}_{\mathrm{p}}}}\right)\Bigg]. \qquad (53)$$

Plugging (40) and (50) – (53) together into (19) we have

$$\mathsf{SINR} \geq \frac{M-K}{\left(K+\frac{1}{\mathsf{SNR}_0}+\sum_{i=1}^{K}\vartheta_{ji}^{(1)}\right)\left(1+\frac{\vartheta_{jk}^{(1)}}{\zeta}+\frac{1}{\mathsf{SNR}_{\mathrm{p}}}\right) + \frac{M-K}{\zeta}\vartheta_{jk}^{(2)} - \sum_{i=1}^{K}\left(\frac{1+\vartheta_{jk}^{(1)}/\zeta+\frac{1}{\mathsf{SNR}_{\mathrm{p}}}}{1+\vartheta_{ji}^{(1)}/\zeta+\frac{1}{\mathsf{SNR}_{\mathrm{p}}}}\right)\left(1+\frac{\vartheta_{ji}^{(2)}}{\zeta}\right)} \qquad (54)$$

This completes the first part of the proof.

---

[12]Hereafter we do not consider the conditioning anymore since, as we will see in a while, we lower bound this term when averaging over the BSs and UEs locations in the outer expectation.

$$\mathbb{E}_{\mathbf{d}}\left\{\frac{1+\frac{\vartheta_{jk}^{(1)}}{\zeta}+\frac{1}{\mathsf{SNR}_{\mathrm{p}}}}{1+\frac{\vartheta_{ji}^{(1)}}{\zeta}+\frac{1}{\mathsf{SNR}_{\mathrm{p}}}}\right\} = \begin{cases} \mathbb{E}_{\mathbf{d}}\{1\}=1 & \text{if } i=k \\ \mathbb{E}_{\mathbf{d}}\left\{1+\frac{\vartheta_{jk}^{(1)}}{\zeta}+\frac{1}{\mathsf{SNR}_{\mathrm{p}}}\right\}\mathbb{E}_{\mathbf{d}}\left\{\frac{1}{1+\frac{\vartheta_{ji}^{(1)}}{\zeta}+\frac{1}{\mathsf{SNR}_{\mathrm{p}}}}\right\}\geq 1 & \text{if } i\neq k \end{cases} \quad (56)$$

$$\mathbb{E}_{\mathbf{d}}\left\{\frac{1+\frac{\vartheta_{jk}^{(1)}}{\zeta}+\frac{1}{\mathsf{SNR}_{\mathrm{p}}}}{1+\frac{\vartheta_{ji}^{(1)}}{\zeta}+\frac{1}{\mathsf{SNR}_{\mathrm{p}}}}\vartheta_{ji}^{(2)}\right\} = \begin{cases} \mathbb{E}_{\mathbf{d}}\left\{\frac{1+\frac{\vartheta_{jk}^{(1)}}{\zeta}+\frac{1}{\mathsf{SNR}_{\mathrm{p}}}}{1+\frac{\vartheta_{jk}^{(1)}}{\zeta}+\frac{1}{\mathsf{SNR}_{\mathrm{p}}}}\vartheta_{jk}^{(2)}\right\}=\mathbb{E}_{\mathbf{d}}\left\{\vartheta_{jk}^{(2)}\right\} & \text{if } i=k \\ \mathbb{E}_{\mathbf{d}}\left\{\frac{\vartheta_{ji}^{(2)}}{1+\frac{\vartheta_{ji}^{(1)}}{\zeta}+\frac{1}{\mathsf{SNR}_{\mathrm{p}}}}\right\}\mathbb{E}_{\mathbf{d}}\left\{1+\frac{\vartheta_{jk}^{(1)}}{\zeta}+\frac{1}{\mathsf{SNR}_{\mathrm{p}}}\right\}\geq \mathbb{E}_{\mathbf{d}}\left\{\vartheta_{ji}^{(2)}\right\} & \text{if } i\neq k \end{cases}$$
(57)

## APPENDIX B — PROOF OF LEMMA 1

Next, a tractable lower bound on the UL average ergodic spectral efficiency of the typical UE is computed, where the expectation is taken with respect to the BSs and UEs locations. To begin with, the Jensen's inequality is applied to move the expectation inside the logarithm and obtain

$$\mathbb{E}_{\mathbf{d}}\left\{\log_2\left(1+\frac{1}{\mathsf{SINR}^{-1}}\right)\right\}\geq \log_2\left(1+\frac{1}{\mathbb{E}_{\mathbf{d}}\{\mathsf{SINR}^{-1}\}}\right)=\log_2\left(1+\underline{\mathsf{SINR}}\right) \quad (55)$$

where we denote with $\underline{\mathsf{SINR}}=\mathbb{E}_{\mathbf{d}}\{\mathsf{SINR}^{-1}\}$. Before proceeding further, let us explicate some of the terms included in (53), that is, (56) and (56), respectively, where in (56) we use the independence between UEs distance realizations for $i\neq k$ together with the Jensen's inequality, while in (57) we lower bound the expectation[13]. From (54), the expectation of $\mathsf{SINR}^{-1}$ can be expanded as in (58).

$$\underline{\mathsf{SINR}} \geq \frac{1}{M-K}\Bigg(\left(K+\frac{1}{\mathsf{SNR}_0}\right)\left(1+\frac{1}{\zeta}\mathbb{E}_{\mathbf{d}}\left\{\vartheta_{jk}^{(1)}\right\}+\frac{1}{\mathsf{SNR}_{\mathrm{p}}}\right)$$
$$+\left(1+\frac{1}{\mathsf{SNR}_{\mathrm{p}}}\right)\sum_{i=1}^{K}\mathbb{E}_{\mathbf{d}}\left\{\vartheta_{ji}^{(1)}\right\}+\frac{1}{\zeta}\sum_{i=1}^{K}\mathbb{E}_{\mathbf{d}}\left\{\vartheta_{jk}^{(1)}\vartheta_{ji}^{(1)}\right\}$$
$$+\frac{M-K}{\zeta}\mathbb{E}_{\mathbf{d}}\left\{\vartheta_{jk}^{(2)}\right\}-K-\frac{1}{\zeta}\sum_{i=1}^{K}\mathbb{E}_{\mathbf{d}}\left\{\vartheta_{ji}^{(2)}\right\}\Bigg). \quad (58)$$

---

[13] Let us consider one term of the sum within $\vartheta_{ji}^{(1)}$ at a time and denote with $x=\beta_{li}^{j}/\beta_{li}^{l}$. Then we have $\mathbb{E}_x\{\frac{x^2}{b+x}\}\geq\frac{(\mathbb{E}_x\{x\})^2}{b+\mathbb{E}_x\{x\}}\geq\frac{\mathbb{E}_x\{x^2\}}{b+\mathbb{E}_x\{x\}}$ by applying Jensen's inequality first (since $b=1+1/\mathsf{SNR}_{\mathrm{p}}>0$) and Holder's inequality at second.

Now, in order to obtain the achievable lower bound in (20) we introduce the following Lemma:

**Lemma 6.** *Assume a multislope path loss model $\beta_{lk}^j(d_{lk}^j)$ as in (2) and $d_{lk}^j \in \Phi_\lambda \setminus \{j\}$ with $d_{lk}^j$ being the distance between UE $k$ in cell $l$ and the BS in cell $j$ (where $\Phi_\lambda \setminus \{j\}$ describes the set of BSs distributed as an H-PPP with density $\lambda$) we have*

$$\mathbb{E}_{\mathbf{d}}\left\{\vartheta_{ji}^{(\kappa)}\right\} = \mathbb{E}_{\mathbf{d}}\left\{\sum_{l \in \Phi_\lambda \setminus \{j\}} \left(\frac{\beta_{li}^j(d_{li}^j)}{\beta_{li}^l(d_{li}^l)}\right)^\kappa\right\} = \mu_\kappa, \tag{59}$$

$$\mathbb{E}_{\mathbf{d}}\left\{\vartheta_{jk}^{(1)}\vartheta_{ji}^{(1)}\right\} = \mathbb{E}_{\mathbf{d}}\left\{\sum_{\substack{n \in \Phi_\lambda \setminus \{j\} \\ l \neq n}} \sum_{l \in \Phi_\lambda \setminus \{j\}} \left(\frac{\beta_{nk}^j(d_{nk}^j)}{\beta_{nk}^n(d_{nk}^n)}\right)\left(\frac{\beta_{li}^j(d_{li}^j)}{\beta_{li}^l(d_{li}^l)}\right)\right\} \leq \mu_1^2 + \mu_2 \tag{60}$$

*where $\mu_\kappa$ for $\kappa = 1, 2$ is denoted in (22).*

*Proof.* We start by considering the BSs distributed in a circular area of finite radius $r$ and wrap around in the radial domain to keep the translation invariance. Considering the closest BS policy association in Section II, which tells us that there are no interfering BSs closer than the one is serving the typical UE $k$ in cell $j$, that is BS $j$, the average number of inter-cell interferers in that ring is $\lambda A_\mathrm{R}$ with $A_\mathrm{R}(r, \lambda) = \pi\left(r^2 - \mathbb{E}_{\mathbf{d}}\left\{(d_{jk}^j)^2\right\}\right) = \pi\left(r^2 - \frac{1}{\pi\lambda}\right)$. Then, (59) – (60) can be written as [13, Appendix B]

$$\mathbb{E}_{\mathbf{d}}\left\{\vartheta_{ji}^{(\kappa)}\right\} = \lambda A_\mathrm{R}(r, \lambda)\, \mathbb{E}_{\mathbf{d}}\left\{\left(\frac{\beta_{lk}^j(d_{lk}^j)}{\beta_{lk}^l(d_{lk}^l)}\right)^\kappa\right\} \tag{61}$$

$$\mathbb{E}_{\mathbf{d}}\{\vartheta_{jk}^{(1)}\vartheta_{ji}^{(1)}\} \leq \lambda A_\mathrm{R}(r, \lambda)\left((\lambda A_\mathrm{R}(r, \lambda) - 1)\mathbb{E}_{\mathbf{d}}^2\left\{\frac{\beta_{lk}^j(d_{lk}^j)}{\beta_{lk}^l(d_{lk}^l)}\right\} + \mathbb{E}_{\mathbf{d}}\left\{\left(\frac{\beta_{lk}^j(d_{lk}^j)}{\beta_{lk}^l(d_{lk}^l)}\right)^2\right\}\right). \tag{62}$$

Therefore, (61) – (62) requires computing the following term

$$\mathbb{E}_{\mathbf{d}}\left\{\left(\frac{\beta_{lk}^j(d_{lk}^j)}{\beta_{lk}^l(d_{lk}^l)}\right)^\kappa\right\} \stackrel{(a)}{=} \mathbb{E}_{\mathbf{d}}\left\{\beta(d_{lk}^l)^{-\kappa}\, \mathbb{E}_{\mathbf{d}}\left\{\beta(d_{lk}^j)^\kappa \mid d_{lk}^l\right\}\right\}$$

$$\stackrel{(b)}{=} \mathbb{E}_{\mathbf{d}}\left\{\beta(d_{lk}^l)^{-\kappa} \int_{d_{lk}^l}^r \beta(x)^\kappa \frac{2x}{r^2 - (d_{lk}^l)^2}\, dx\right\} \tag{63}$$

$$\stackrel{(c)}{=} \sum_{n=1}^N \left(\int_{R_{n-1}}^{R_n} \frac{2\beta_n(y)^{-\kappa}}{r^2 - y^2}\left(\int_y^r x\, \beta(x)^\kappa\, dx\right) f_d(y)\, dy\right)$$

for $\kappa = \{1, 2\}$ and with $\beta_n$ as the path loss related to the interval $[R_{n-1}, R_n)$, where in $(a)$ we use the theorem of total expectation conditioning over $d_{lk}^l$, $(b)$ is due to the fact that there are no interfering BSs closer than the $j$th (i.e., $d_{lk}^j \geq d_{lk}^l$) and change of variable to polar coordinates, while in $(c)$ we use the multislope pathloss model in (2) $y \in [R_{n-1}, R_n)$, $n = 1, \ldots, N$ and denote with $f_d(d_{jk}^j)$ the probability density function of the distance from the typical UE $k$ to its serving BS $j$. By using (2), we rewrite the term within the inner brackets in (63) as follows

$$\int_y^r x\beta(x)^\kappa \, dx = \int_y^{R_n} x\beta_n(x)^\kappa \, dx + \sum_{i=n+1}^N \int_{R_{i-1}}^{R_i} x\beta_i(x)^\kappa \, dx \tag{64}$$

where we have used that $x \geq y$ for any fixed index $n$ of the summation in (63). Then, using (2) into (64) we obtain

$$\int_y^r x\beta(x)^\kappa \, dx = \Upsilon_n^\kappa \frac{y^{2-\kappa\alpha_n} - R_n^{2-\kappa\alpha_n}}{\kappa\alpha_n - 2} + \sum_{i=n+1}^N \Upsilon_i^\kappa \frac{R_{i-1}^{2-\kappa\alpha_i} - R_i^{2-\kappa\alpha_i}}{\kappa\alpha_i - 2}. \tag{65}$$

Plugging (65) into (63), after some rearrangements we obtain

$$\mathbb{E}_\mathbf{d}\left\{\left(\frac{\beta_{lk}^j(d_{lk}^j)}{\beta_{lk}^l(d_{lk}^l)}\right)^\kappa\right\} = 2\sum_{n=1}^N \int_{R_{n-1}}^{R_n} \left(\frac{y^2}{(\kappa\alpha_n - 2)(r^2 - y^2)} + \frac{y^{\kappa\alpha_n} c_n(\kappa)}{(r^2 - y^2)}\right) f_d(y) \, dy \tag{66}$$

where $c_n(\kappa) = \sum_{i=n+1}^N \left(\frac{\Upsilon_i}{\Upsilon_n}\right)^\kappa \frac{R_{i-1}^{2-\kappa\alpha_i} - R_i^{2-\kappa\alpha_i}}{\kappa\alpha_i - 2} - \frac{R_n^{2-\kappa\alpha_n}}{\kappa\alpha_n - 2}$.

To account for the wrap around in $\mathbb{R}^2$, we let $r \to \infty$ and obtain[14]

$$\lambda A(r, \lambda) \mathbb{E}_\mathbf{d}\left\{\left(\frac{\beta_{lk}^j(d_{lk}^j)}{\beta_{lk}^l(d_{lk}^l)}\right)^\kappa\right\} \to 2\lambda\pi \sum_{n=1}^N \left(\frac{1}{\kappa\alpha_n - 2} \int_{R_{n-1}}^{R_n} y^2 f_d(y) \, dy + c_n(\kappa) \int_{R_{n-1}}^{R_n} y^{\kappa\alpha_n} f_d(y) \, dy\right). \tag{67}$$

Finally, by using Corollary 3 the two integrals in (66) can be computed in closed-form as (after simple calculus)

$$\int_{R_{n-1}}^{R_n} y^2 f_d(y) \, dy = \frac{\Gamma(2; \pi\lambda R_{n-1}^2) - \Gamma(2; \pi\lambda R_n^2)}{\pi\lambda} \tag{68}$$

$$\int_{R_{n-1}}^{R_n} y^{\kappa\alpha_n} f_d(y) \, dy = \frac{\Gamma\left(\frac{2+\kappa\alpha_n}{2}; \pi\lambda R_{n-1}^2\right) - \Gamma\left(\frac{2+\kappa\alpha_n}{2}; \pi\lambda R_n^2\right)}{(\pi\lambda)^{\frac{\kappa\alpha_n}{2}}}. \tag{69}$$

Plugging (67) – (69) into (61) – (62) leads to (59) – (60), which completes the proof. □

---

[14] Notice that the function within the integral is a fractional polynomial of second order degree that converges to a bounded real-valued limit as $r \to \infty$. For this reason, the bounded convergence theorem conditions are satisfied and that operation is allowed.

## APPENDIX C — PROOF OF LEMMA 5

We start by noticing that the mapping applied at each iteration, to compute the next point of the algorithm, follows the gradient descent rule, which provides a stationary point with no increase of the objective value. In [32, Theorem 3.2], convergence of a gradient descent-based alternating method is proved given that the objective $\underline{\text{EE}}(\bar{c}, K)$ is pseudo-concave in each variable and the level set $\mathcal{S} = \{(\bar{c}, K) \in \mathbb{R}^2 : \underline{\text{EE}}(\bar{c}, K) \leq \underline{\text{EE}}(\bar{c}_0, K_0)\}$, with respect to the the initial point of the algorithm, is compact. The latter can be easily proved by noticing that $\bar{c}$ and $K$ are defined over a box (compact and closed domain) and the objective function is bounded, which is true for any $P_{\text{FIX}} > 0$. In particular, under these conditions, the alternating optimization method introduced in Lemma 5 returns a sequence $\{\bar{c}_k, K_k\}$ that converges to a global maximizer of $\underline{\text{EE}}$ on $\mathbb{R}^2$. Thus, convergence is achieved whenever pseudo-concavity holds component-wise for both $\underline{\text{EE}}(\bar{c})$ and $\underline{\text{EE}}_\infty(K)$ in (30) and (39), respectively. To this end, in [33, Proposition 2.9] it is shown that fractional objectives in the form of $r = f/g$ defined over a convex set $\mathcal{X} \subseteq \mathbb{R}^n$ with $n \geq 1$ enjoy pseudo-concave properties when $f : \mathcal{X} \to \mathbb{R}$ is non-negative, differentiable, and concave, while $g : \mathcal{X} \to \mathbb{R}$ is differentiable and convex (if $g$ is affine, the non-negativity of $f$ can be relaxed). In particular, the former objective can be rewritten as $r(\bar{c}) = \underline{\text{EE}}(\bar{c})$ with $f(\bar{c}) = p_1 \bar{c} + p_0$ and $g(\bar{c}) = q_2 \bar{c}^2 + q_1 \bar{c} + q_0$ with some parameters $\{p_i, q_i\}$ that can be easily related to the parameters $\{a_i\}$ illustrated in Table III. Here, $f(\bar{c})$ is affine and non-negative for $\bar{c} \geq -\frac{p_0}{p_1} = \frac{a_1 + a_3}{a_2 - a_0}$, which is equivalent to assume $\gamma \leq \frac{\tau_c}{\mu_2 K}$ (true in general as seen below (30)) whereas $g(\bar{c})$ is convex since $q_2 = a_2 a_4$ that are both positive quantities; see Table III. For the latter objective $r(K) = \underline{\text{EE}}_\infty(K)$, we have that $f(K) = -m_3 K^2 + m_2 K$ is concave for $m_3 = b_0 > 0$ (see Table III), while $g(K) = n_2 K + n_1$ is affine over $K \in \mathbb{R}$. This proves the convergence of Algorithm 1.